\documentclass{nature}

\usepackage{fixltx2e}
\usepackage{amsmath}
\usepackage{amssymb}
\usepackage{epsfig}
\usepackage{epstopdf}
\usepackage{rotating}
\usepackage{caption,subcaption}
\usepackage{color}
\usepackage{colortbl}
\usepackage{graphicx}
\usepackage{subfig}
\usepackage{ifthen}
\usepackage{alltt}
\usepackage{float}
\usepackage{fancyhdr}
\usepackage{verbatim}
\usepackage{moreverb}
\usepackage{cite}
\usepackage{multirow}
\usepackage{bigstrut}
\newcommand{\onlinecite}[1]{\hspace{-1 ex} \nocite{#1}\citenum{#1}} 

\title{Functional metamirrors}

\author{V.~S.~Asadchy$^{1,2}$, Y.~Ra'di$^1$, J.~Vehmas$^1$ \& S.~A.~Tretyakov$^1$}

\begin{document}

\maketitle

\begin{affiliations}
 \item Department of Radio Science and Engineering, Aalto University, P.~O.~Box~13000, FI-00076 Aalto, Finland
 \item Department of General Physics, Gomel State University, 246019, Belarus
\end{affiliations}

\begin{abstract}


Conventional mirrors obey Snell's reflection law: a plane wave is reflected as a plane wave, at the same angle. To engineer spatial distributions of fields reflected from a mirror, one can either shape the reflector (for example, creating a parabolic reflector) or position some phase-correcting elements on top of a mirror surface (for example, designing a reflectarray antenna). Here we show, both theoretically and experimentally, that full-power reflection with general control over reflected wave phase is possible with a single-layer array of deeply sub-wavelength inclusions. These proposed artificial surfaces, metamirrors, provide various functions of shaped or nonuniform reflectors without utilizing any mirror. This can be achieved only if the forward and backward scattering of the inclusions in the array can be engineered independently, and we prove that it is possible using electrically and magnetically polarizable inclusions. 
The proposed sub-wavelength inclusions possess desired reflecting properties at the operational frequency band, while at other frequencies the array is practically transparent. The metamirror concept  leads to a variety of applications over the entire electromagnetic spectrum, such as optically transparent focusing antennas for satellites, multi-frequency reflector antennas for radio astronomy, low-profile conformal antennas for telecommunications, and nano-reflectarray antennas for integrated optics.

\end{abstract}

Conventional mirrors, known  since the dawn of civilization\cite{ancient}, obey the simple law of reflection: the reflection angle is equal to the incidence angle. This follows from the fact that the total tangential electric field at the ideal mirror surface is zero, thus, the phase of the electric field in the reflected wave is the opposite to that in the incident wave. If a reflector can be engineered to enable general control over the reflection phase, it is possible to change the direction of the reflected waves at will\cite{capasso}.  Developments in the field of antennas enabled creation of layers with any desired phase of reflection at microwaves.   Conceptually, these devices are conventional mirrors, modified by some additional phase-shifting elements positioned close to fully reflecting surfaces.   Such artificial layers, in particular conventional reflectarrays\cite{reflectarray} and high-impedance surfaces\cite{HIS}, are realized as arrays of resonant metal patches over a metal ground plane. All the patches of the array usually have identical shape but different dimensions, so that the resonance frequency of an individual patch varies to ensure the desired variations of the reflection phase over the array surface. Controlling the phase variation spanning a $2\pi$ range allows one to  arbitrarily tune the shape and orientation of the reflected wavefront. Since the conventional reflectarrays incorporate a metal ground plane, the transmission through them is completely blocked and reflection amplitude can be very high if low-loss materials are used. On the other hand, the presence of a metal ground plane forbids transmission at all practically interesting frequencies and limits the application possibilities. 

If it will become possible to create single layers of small particles which fully reflect incident electromagnetic waves and allow full control over reflection phase, all the limitations due to the presence of a continuous conventional mirror will be removed. 
Moreover, the absence of a ground-plane reflector makes it possible to independently control response to incident waves impinging on the structure from the opposite sides. 
We expect that such structures could find numerous applications in communication and information processing, for example, as optically transparent microwave reflectarrays, band-stop filters for wave trapping, and asymmetric frequency-selective mirrors with desired reflection phases.

Clearly, such sub-wavelength structural layers possess electromagnetic properties which are not available in natural materials, and potential realizations require the use of artificial materials, called metamaterials. Recent advances in metamaterials\cite{cloaking,negative} have opened new routes towards engineering artificial structures with unprecedented control over their interaction with incident radiation. `Metamaterials', derived from Greek `beyond (conventional) materials', are macroscopic composites having a sub-wavelength architecture engineered to provide desired electromagnetic properties. 
The last few years have witnessed remarkable progress
in the development of single-layer planar metamaterials having deeply sub-wavelength thicknesses. Such two-dimensional composites, so-called metasurfaces\cite{capasso}, have demonstrated capability to arbitrarily tailor the reflected and transmitted wavefronts. This becomes possible when the inclusions of a metasurface are designed in such a way that there is a gradient of phase discontinuity over the surface. This well-known phenomenon from antenna array theory has been applied to geometrical optics and yielded the generalized laws of reflection and refraction\cite{snell}. Anomalous reflection and refraction provide great flexibility in shaping reflected and transmitted beams through thin planar surfaces.

A majority of recent works on metamaterials for field control focus on the control of transmission. Electrically thin metasurfaces supporting electric surface currents have been proposed for wavefront engineering of transmission with the phase variation spanning a full 2$\pi$ range\cite{snell,shalaev1,shalaev2,babinet,mosall1}. The main drawback of these known designs is that the metasurfaces tailor wavefronts of transmitted waves only for cross polarization (polarized in the direction perpendicular to the polarization of incident waves). Moreover, the structures are polarization-selective and possess low efficiency with a theoretical upper bound of 25\%\cite{alu}. The efficiency can be enhanced up to 75\% by adding metal cut-wire arrays\cite{terahertz} that filter out the undesired polarization, but this significantly increases the thickness and absorption loss level.
Alternative approach for wavefront manipulation in transmission at optical frequencies relies on excitation of localized plasmon resonances in metallic nanostructures\cite{LSP1,LSP2}. However, this approach does not provide full phase control of waves covering only $\pi$ phase range, which limits its application potentials.
In the last few years, complete and efficient control of electromagnetic wavefronts in transmission has been achieved in metasurfaces that possess both electric and magnetic responses. This was recently demonstrated at microwave\cite{grbic1} and optical frequencies\cite{alu,grbic2}. These metasurfaces efficiently operate with co-polarized waves (without polarization alteration) and support phase variation of the transmission spanning a full 2$\pi$ range.

While metasurfaces tailoring wavefronts in transmission usually consist of sub-wavelength inclusions and are transparent outside of the operating frequency band, most metasurfaces manipulating reflection are metal-backed and detectable over the entire frequency range\cite{sun,bozh1,bozh2,mosall2,nanotechnology}. To the best of our knowledge, the only work on phase-controlling reflecting structures without a ground plane utilizes a multi-layer electrically thick structure\cite{weide}.
The difficulties in extending the well-understood techniques for phase control in transmission to control of reflection phase  follow from the fact that the 
physical phenomena behind manipulation of wavefronts in transmission and reflection are crucially different. Transmission wavefront control can be accomplished by using an array of various so-called Huygens' elements (zero-backward scattering inclusions) which scatter waves with adjustable phase in the forward direction. To control the wavefront of reflection, the inclusions in the array should be engineered in such a way that they re-radiate waves in the backward direction with different phases, while in the forward direction they scatter waves with the same phase, opposite to that of the incident plane wave. 

Here, we introduce and experimentally validate a  new  concept of non-uniform \emph{metamirrors}, ultra-thin engineered structures providing full control of reflected wavefronts. Due to the lack of a ground plane and deeply sub-wavelength dimensions of the inclusions, the designed metamirrors are practically invisible at frequencies outside of the operational band.
We utilize the theoretical idea of uniform full-reflection sub-wavelength metasurfaces\cite{mirror1} and develop it for realization of non-uniform phase distributions of reflected plane waves.
The proposed metamirrors are formed by single planar arrays of specifically shaped resonant bianisotropic inclusions\cite{serdukov,mirror1} possessing both electric and magnetic responses, as well as magnetoelectric coupling. We show that utilizing general reciprocal bianisotropic properties of the inclusions we can achieve complete control of reflection phase variations independently for the two sides of the metamirror. This feature opens new opportunities for engineering ultra-thin asymmetric two-sided reflectarrays.
With the use of low-loss materials, the proposed design can approach 100$\%$ efficiency without polarization conversion.
To design and optimize metamirrors, we use the metasurface modeling method based on polarizabilities of bianisotropic unit cell\cite{mirror1} and demonstrate performance of two metamirror designs with specific wavefront shaping properties. 
Conceptual experimental measurements have been carried out at microwaves, although the metamirrors concept can be applied over the entire electromagnetic spectrum.

\section*{Results}
\subsection{Metamirror concept.}
\begin{figure*}[h]
\centering
\begin{subfigure}{0.28\columnwidth}
  \centering
  \includegraphics[width=\columnwidth]{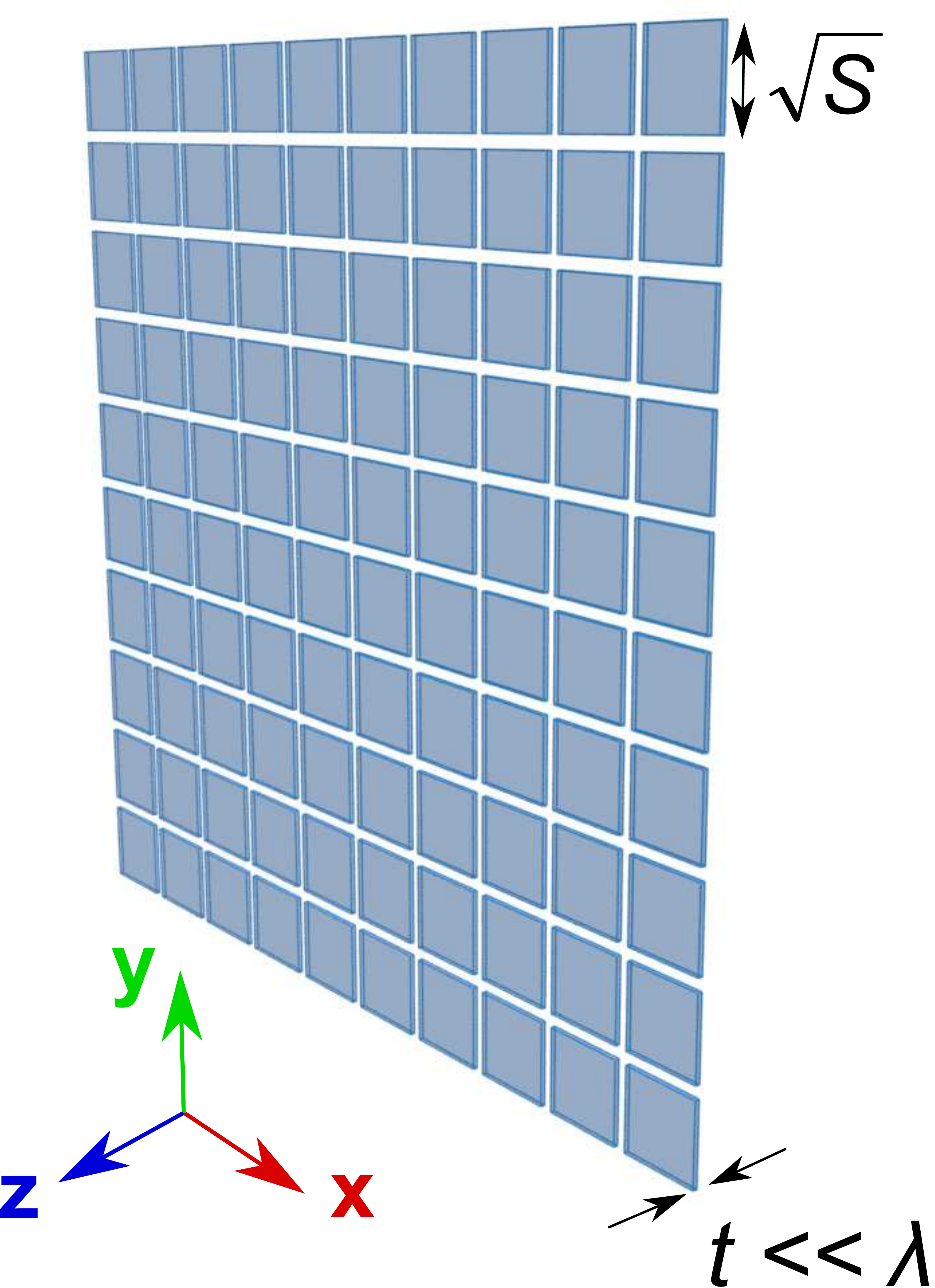}
  \caption{}
  \label{fig:fig1a}
\end{subfigure}\hspace{2mm}
\begin{subfigure}{0.28\columnwidth}
  \centering
  \includegraphics[width=\columnwidth]{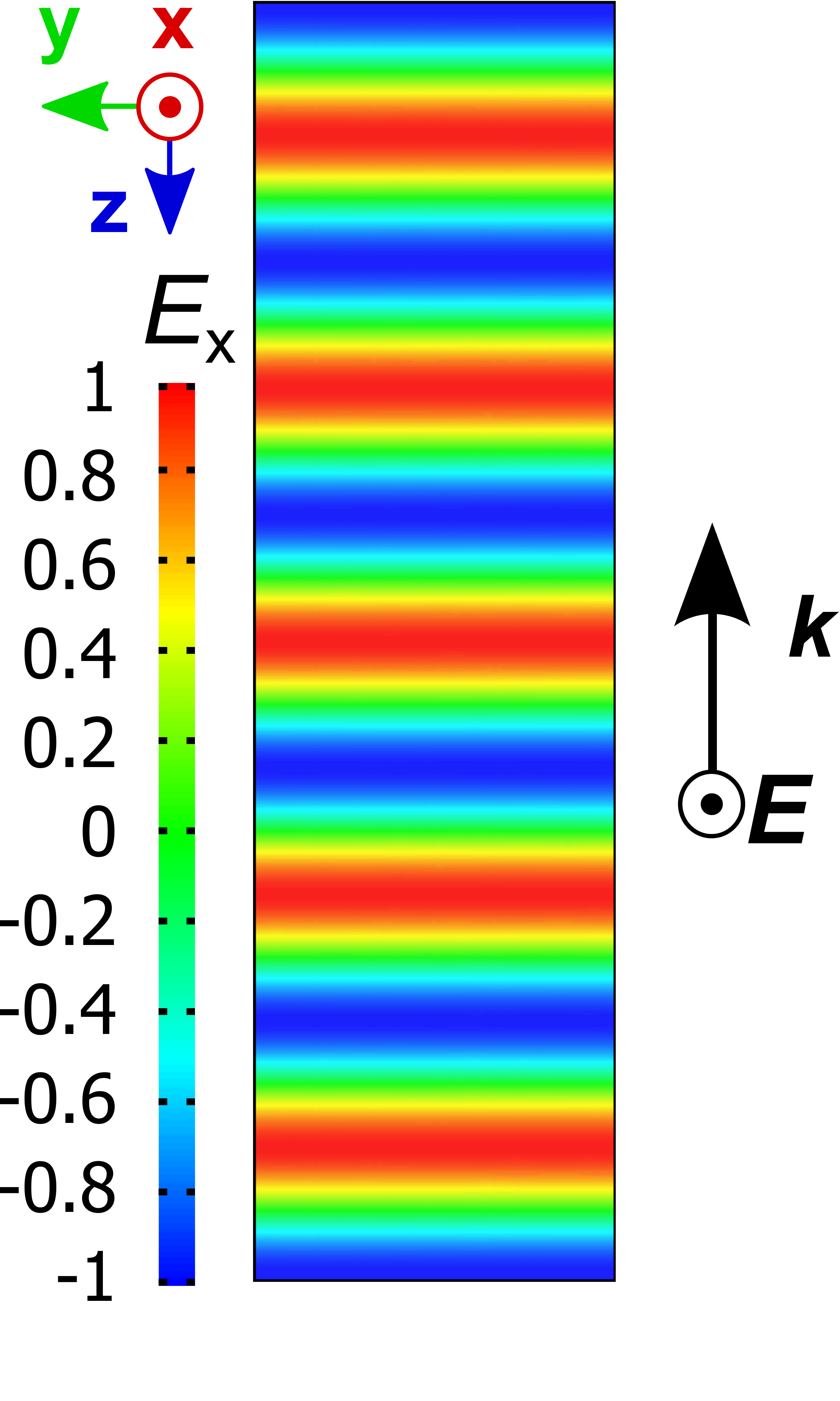}
  \caption{}
  \label{fig:fig1b}
\end{subfigure}\hspace{2mm}
\begin{subfigure}{0.28\columnwidth}
  \centering
  \includegraphics[width=\columnwidth]{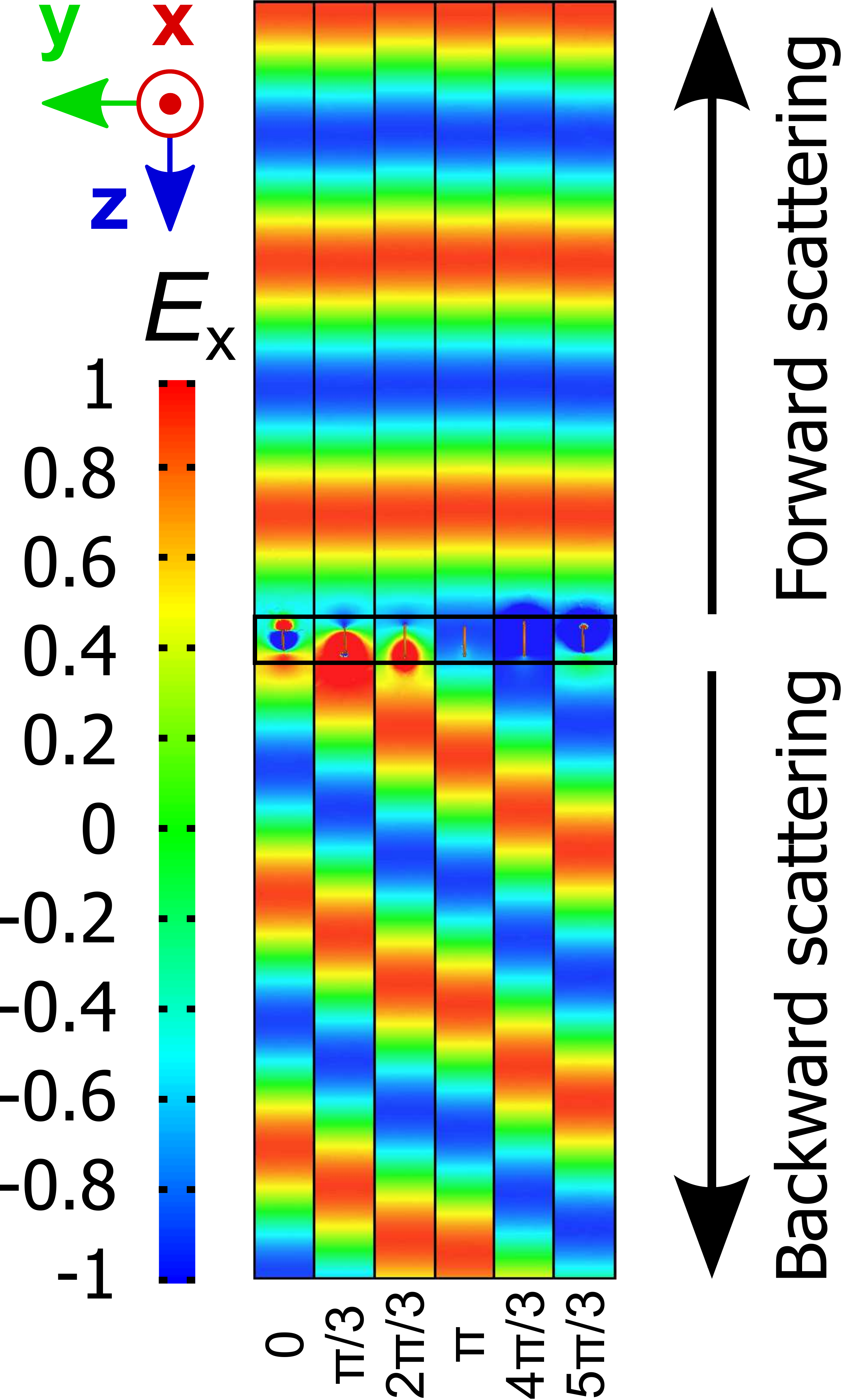}
  \caption{}
  \label{fig:fig1c}
\end{subfigure}%
\caption{Asymmetrical scattering properties of a metamirror. (a) An illustration of a generic, optically thin metamirror. The results of full-wave simulations of the individual inclusions of a metamirror which totally reflects normally incident waves at an angle $\theta=45^{\circ}$ from the normal. (b) Field distribution of an incident wave normally impinging on the metamirror surface. (c) Distribution of the co-polarized electric field of the scattered waves from each inclusion. The forward scattered waves from the individual inclusions have identical phases opposite to the phase of the incident wave, which yields zero transmission. In the backward direction the inclusions radiate with discrete phase shifts from 0 to $5\pi/3$, clearly revealing anomalous reflection.}
\label{fig:fig1}
\end{figure*}

Consider a metasurface consisting of a single planar periodic array of identical sub-wavelength inclusions polarizable both electrically and magnetically. An incident plane wave impinges on the array normally to its surface along $z$-direction (see Fig.~\ref{fig:fig1a}). 
Since the array period is small compared with the wavelength, the electric and magnetic moments induced in the inclusions can be modelled as surface-averaged electric and magnetic currents. The induced currents radiate secondary plane waves in the backward and forward directions. As shown in ref.~\onlinecite{mirror1}, it is convenient to describe the properties of a metasurface in terms of the polarizabilities $\alpha$ of its unit cell of the area $S$.
The electric fields of the backward and forward scattered plane waves from the metasurface illuminated by an incident plane wave are given by
\begin{equation}
\begin{array}{c}\displaystyle
\mathbf{E}_{\rm
back}=-\frac{j\omega}{2S}\left(\alpha_{\rm e}\pm 2\alpha_{\rm me}-
\alpha_{\rm m}\right)\mathbf{E}_{\rm inc},\qquad
\mathbf{E}_{\rm
forw}=-\frac{j\omega}{2S}\left(\alpha_{\rm e}+
\alpha_{\rm m}\right)\mathbf{E}_{\rm inc},
\end{array}
\label{eq:1}
\end{equation}
\noindent
where $\omega$ is the angular frequency, $j$ is the imaginary unit,
$\alpha_{\rm e}$, $\alpha_{\rm m}$, and $\alpha_{\rm me}$ are, respectively, the electric, magnetic, and so-called magnetoelectric polarizabilities of the unit cell, normalized to the impedance of free space.
The upper and lower signs in the equation correspond to two different cases when the metamirror is illuminated from the $+z$ and $-z$ directions, respectively.
The polarizability quantities characterize the electric and magnetic responses of the inclusions as well as their mutual interactions in the array. The magnetoelectric polarizability $\alpha_{\rm me}$ measures the capability of the inclusion to acquire electric polarization under the influence of an external magnetic field. Such property, inherent in bianisotropic inclusions\cite{serdukov}, enables an additional freedom in metasurface engineering. 

Zero transmission through the metasurface is achieved when the inclusions collectively radiate a secondary wave in the forward direction which destructively interferes with the incident wave $\displaystyle \mathbf{E}_{\rm forw}=-\mathbf{E}_{\rm inc}$. To ensure full reflection with any desired phase $\phi$, the backscattered wave must be such that $\displaystyle \mathbf{E}_{\rm back}=e^{j \phi}\mathbf{E}_{\rm inc}$. We use these two conditions to determine the necessary polarizabilities of the unit cells and subsequently design the inclusions with the desired properties. 

It should be stressed here that the complete phase control of reflection can be achieved by exploiting an array of anisotropic particles with the zero $\alpha_{\rm me}$ and non-zero $\alpha_{\rm e}$ and $\alpha_{\rm m}$ polarizabilities such as simple electric and magnetic dipoles. However, the design of such an array becomes very challenging since the inclusions of these two types must operate in a metamirror at non-resonant frequencies\cite{mirror1} and be adjusted very precisely taking into account their mutual interaction. In order to relax the design requirements, we utilize only one type of inclusions with the additional non-zero magnetoelectric polarizability $\alpha_{\rm me}$. Such inclusions, in addition to considerably smaller sizes, operate at the resonance frequency and enable asymmetrical reflection properties of the metamirror from its two sides.
Generalized relations (\ref{eq:1}) are written for the case of a metamirror with uniaxial symmetry (rotation of the metasurface around the normal does not affect its electromagnetic response). Therefore, our metamirrors operate identically for any polarization of the incident~wave.

In order to efficiently manipulate wavefronts of reflection from metamirrors, we adjust the phase of reflection from each inclusion individually, preserving the unity value for the reflected-wave amplitude. This requires that all the inclusions radiate waves in the forward direction with the identical phase (opposite to the phase of the incident wave) and in the backward direction the phase of the radiation varies according to a specific distribution along the surface. This feature dramatically distinguishes the metamirror concept from the designs utilizing a ground plane\cite{sun,bozh1,bozh2,mosall2,nanotechnology}. It can be seen from equations (\ref{eq:1}) that different scattering properties in the backward and forward directions can be accomplished only if the inclusions possess both electric and magnetic responses.
Such asymmetrical scattering is illustrated in Fig.~\ref{fig:fig1}, where an array of inclusions radiates waves in the backward direction with a linearly varying phase. 
This metamirror totally reflects incident waves impinging normally on its surface (Fig.~\ref{fig:fig1b}) at an angle $\theta=45^{\circ}$ from the normal (Fig.~\ref{fig:fig1c}). Generally, metamirrors provide an efficient route to achieve an arbitrary reflection phase distribution in the full-reflection regime. The concept of metamirrors can be further extended to the additional control of the reflection amplitude by introducing the absorption losses\cite{absorption} or non-reciprocity\cite{transparency} in the structure.

\subsection{Anomalous reflection.}
To demonstrate the potential and flexibility of the proposed metamirror concept, we consider two example structures: a metamirror anomalously reflecting normally incident waves and a reflecting \emph{metalens} (a single-layer focusing sheet). Although the proposed concept is generic and can be applied over the entire electromagnetic spectrum, we design and experimentally measure structures operating in the microwave frequency band (in the G-band). The designed prototypes can be subsequently pushed to infrared and even optical range by downscaling the sizes of the structural inclusions, modifying their geometry, and taking into consideration the properties of materials at different frequencies. In our design we content ourselves with the case of metamirrors operating with incident waves of a single linear polarization. This clarifies the engineering procedure and simplifies the experimental realization. Polarization-insensitive regime in the metamirrors can be achieved by the use of inclusions possessing uniaxial symmetry\cite{mirror1}, which can be realized simply by adding a second particle (identical in shape and size but oriented orthogonally to first particle) into each unit cell.

\begin{figure*}[h]
\centering
\begin{subfigure}{0.31\columnwidth}
  \centering
  \includegraphics[width=\columnwidth]{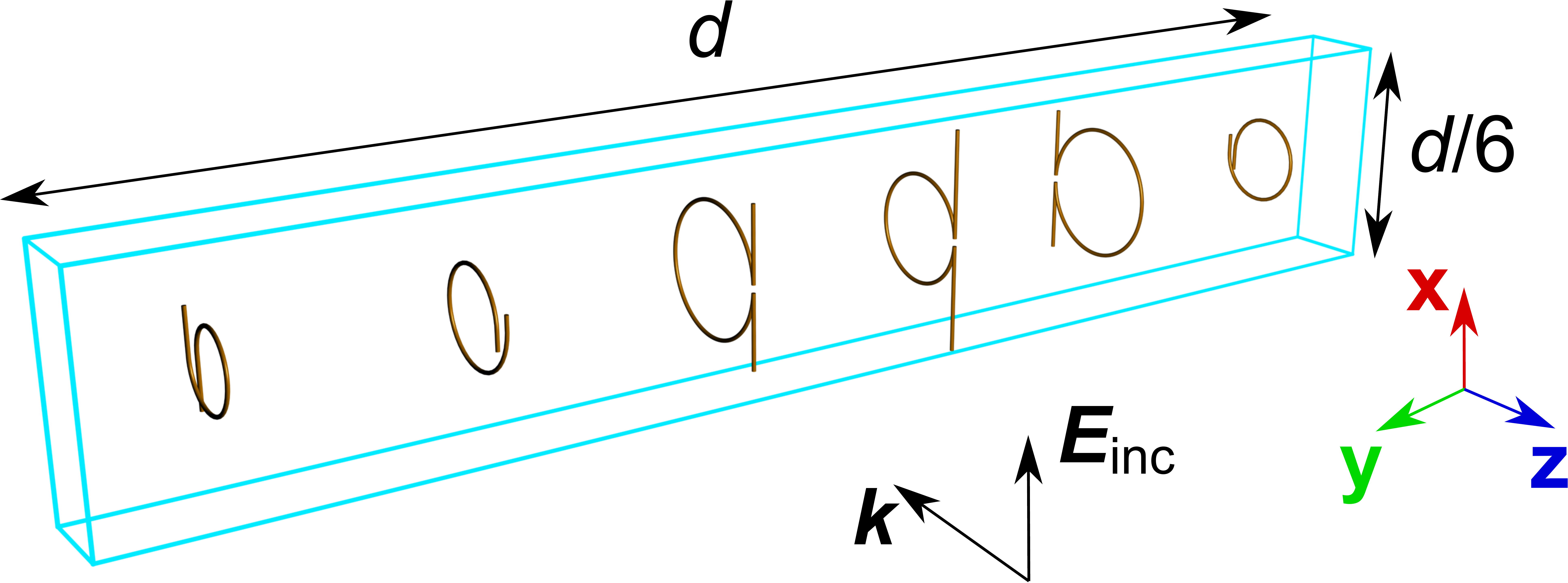}
  \caption{}
  \label{fig:fig2a}
\end{subfigure}\hspace{2mm}
\begin{subfigure}{0.31\columnwidth}
  \centering
  \includegraphics[width=\columnwidth]{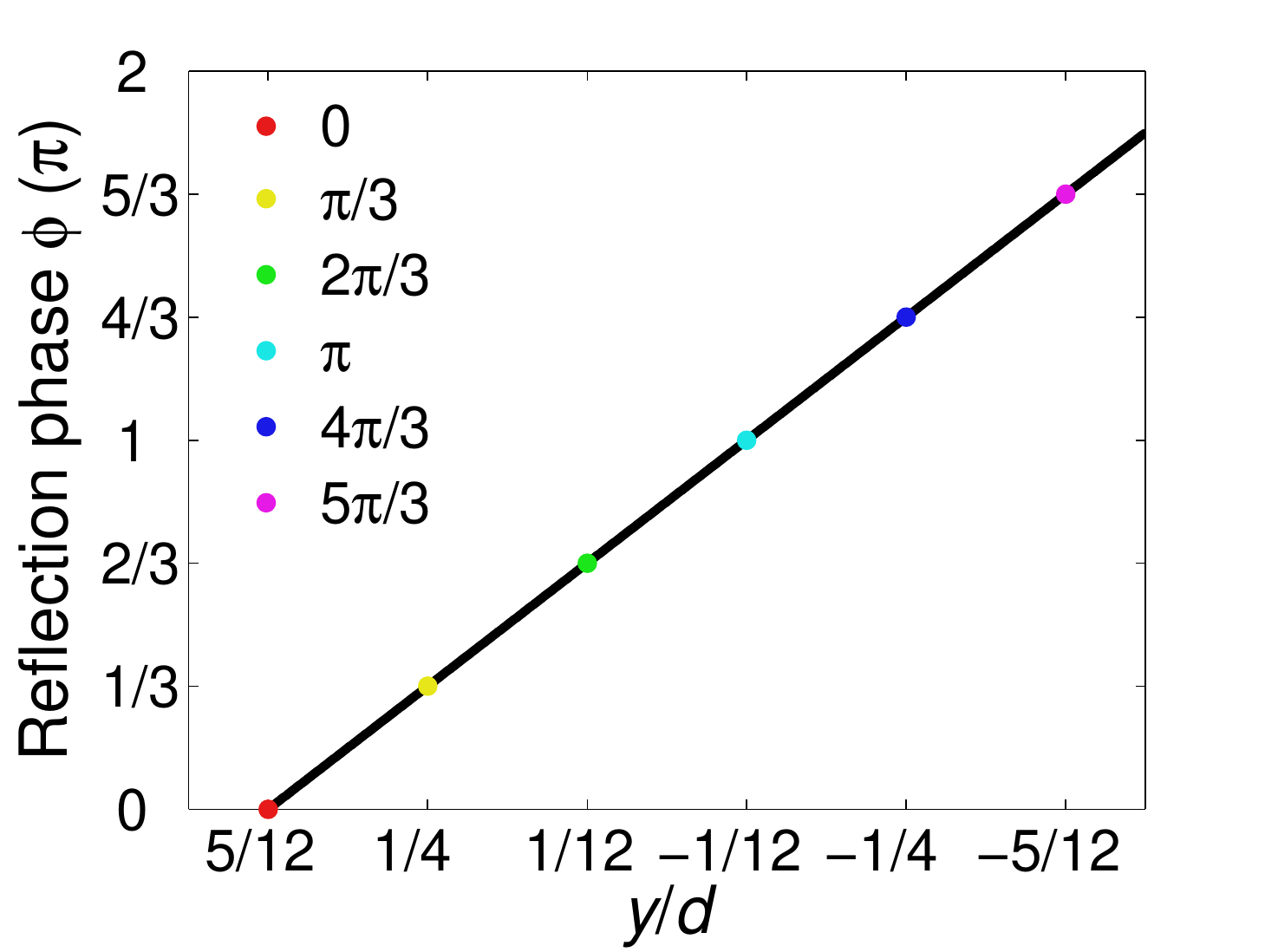}
  \caption{}
  \label{fig:fig2b}
\end{subfigure}\hspace{2mm}
\begin{subfigure}{0.31\columnwidth}
  \centering
  \includegraphics[width=\columnwidth]{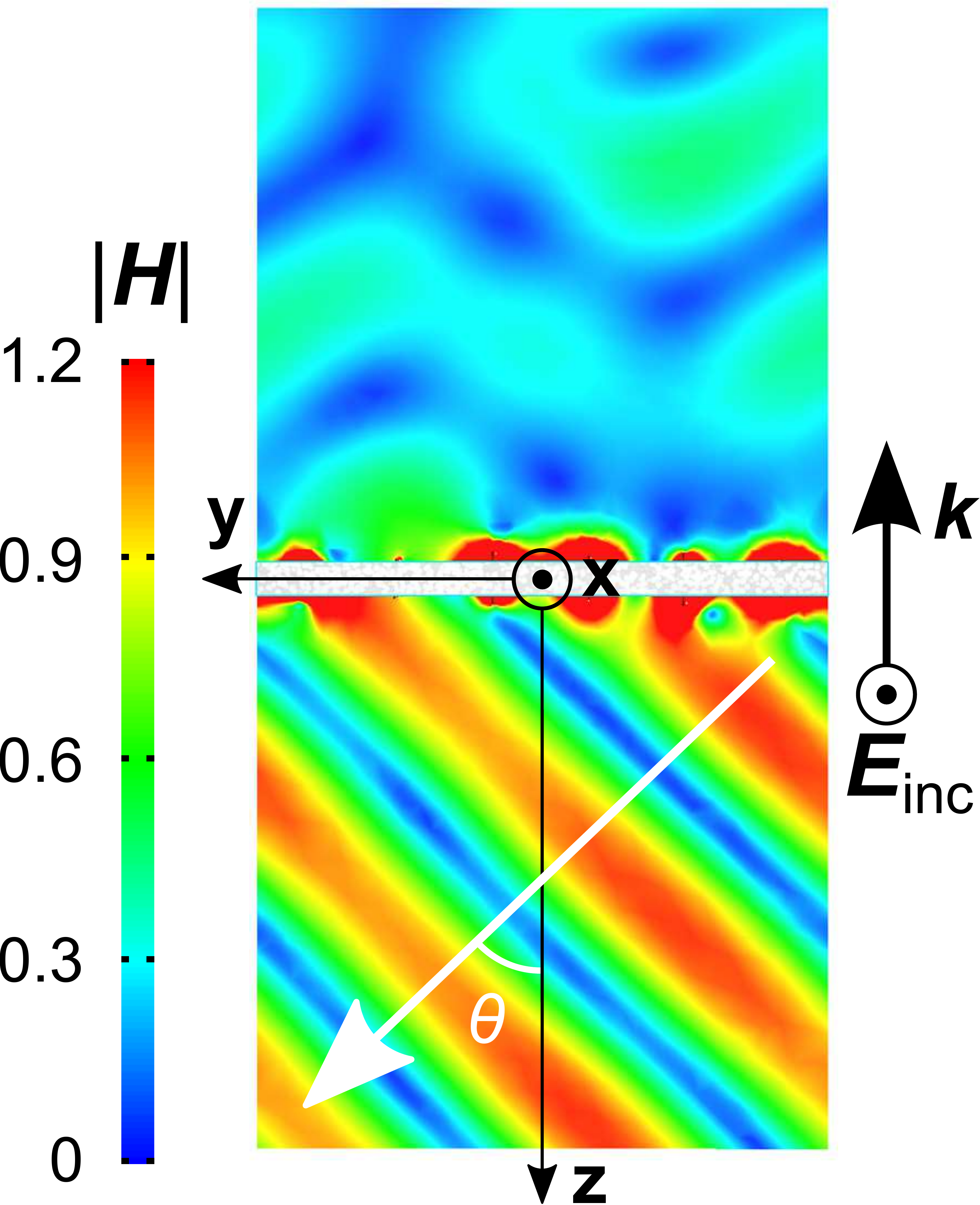}
  \caption{}
  \label{fig:fig2c}
\end{subfigure}%
\caption{Anomalous reflection of normally incident plane waves. (a) The period of a metamirror consisting of 6 sub-wavelength copper inclusions which provide a linear phase variation of the reflection spanning a $2\pi$ range. The blue box denotes the dielectric substrate. (b) The solid curve shows the required phase variations at the metamirror surface within one period $d$. The dots of different colors show the phase that is provided by the different designed inclusions. (c) Magnetic field distribution (normalized to the magnetic field of the incident wave) of the transmitted (the $-z$~half-space) and reflected (the $+z$~half-space) waves.}
\label{fig:fig2}
\end{figure*}

First, we design  a metamirror to efficiently reflect normally incident plane waves to an arbitrarily chosen angle $\theta=45^{\circ}$ from the normal. To realize this functionality, a linear phase variation of reflection along the metamirror is required. Based on the principle of phased arrays, the reflected wavefront is deflected to an angle $\theta$ if the metasurface provides a linear phase variation spanning the 2$\pi$ range with the periodicity $d=\lambda/\sin\theta$, where $\lambda$ is the operating wavelength. 
In our design we use copper wire inclusions embedded in Rohacell-51HF ($\epsilon_{\rm r}=1.065$, $\rm tan \delta=0.0008$) foam of 5~mm thickness (see Fig.~\ref{fig:fig2a}). The dielectric substrate is needed only for mechanical support of small conductive inclusions.
Each period $d$ is discretized into 6 unit cells consisting of different sub-wavelength inclusions providing discrete phase shifts shown in Fig.~\ref{fig:fig2b}. The use of sub-wavelength inclusions ensures homogeneous (in the sense that the averaged electromagnetic response varies smoothly over the structure) phase variation along the interface. All the inclusions have non-zero magnetoelectric polarizability $\alpha_{\rm me}$ due to the fact that they incorporate electrically polarizable  straight wires connected to magnetically polarizable wire loops. 
The required polarizabilities $\alpha_{\rm e}$, $\alpha_{\rm m}$, and $\alpha_{\rm me}$ of the inclusions were found based on equations (\ref{eq:1}). In order to determine the shape and dimensions of practically realizable inclusions with the required polarizabilities, we utilize an approach based on scattering cross sections\cite{polarizability}. All the dimensions of the inclusions and description of their electromagnetic polarizabity can be found in Supplementary Note~1, Supplementary Fig.~S1, and Supplementary Table~S1.

The simulated scattering properties of the inclusions of the metamirror are shown in Fig.~\ref{fig:fig1c}. The operating frequency of 5~GHz is chosen. At this frequency the periodicity along the $y$-axis is $d=\lambda/\sin45^{\circ}=84.9$~mm (impact of the substrate has been neglected). Along the $x$-axis the unit cells are positioned with the periodicity $d/6=14.1$~mm. An incident plane wave impinges on the metamirror from the $+z$-direction with the electric field parallel to the $x$-axis.
Figure~\ref{fig:fig2c} illustrates simulated magnetic field distribution of the reflected (backward scattered wave) and transmitted waves (interference of the incident and forward scattered waves).  According to the design target, the wavefront of the reflection is planar and deflected at an angle of $45^{\circ}$ with respect to the normal. 
Only 8$\%$ of the incident power is transmitted and 6$\%$ is absorbed by the material of the inclusions. The high reflectance of 86$\%$ and the pure wavefront rotation confirm remarkable functionality of the metamirror. The reflection level can be further improved by optimizing the inclusions taking into account the mutual interaction of inclusions of different types. The half-power bandwidth of the proposed metamirror is 0.25~GHz or 5\%, which can be considered quite broad taking into account the sub-wavelength thickness $t=\lambda/7$ of the designed structure. The frequency stability of the metamirror performance is shown in Supplementary Fig.~S2.
It should be stressed that the proposed metamirror based on the wire inclusions is only a conceptual prototype and can be improved by using inclusions of other types\cite{other1,other2} with the properties dictated by equations (\ref{eq:1}). The functionality of the metamirror for incident waves from the $-z$-direction is presented in Supplementary Fig.~S3.

\subsection{Focusing at sub-wavelength distances.}
Another example demonstrating the universality of non-uniform metamirrors is focusing metasurfaces showing extremely strong wave-gathering ability. 
The metamirror concept opens a new route towards engineering a conceptually new kind of lenses: one consisting of an ultimately thin single layer of sub-wavelength inclusions providing near-diffraction-limit focusing of electromagnetic energy at a chosen point. Sub-wavelength sizes of the inclusions which radiate as dipoles provide homogeneous phase variation along the surface and enable focusing at extremely short distances.
Since the metalens is planar, it does not suffer from spherical aberrations inherent in conventional focusing reflectors.

To demonstrate the lensing effect, we have designed a single-layer
metamirror composed of 6 concentric arrays of sub-wavelength inclusions with the dimensions described in Supplementary Table~S2 (see Fig.~\ref{fig:fig3a}). 
\begin{figure*}[h]
\centering
\begin{subfigure}{0.31\columnwidth}
  \centering
  \includegraphics[width=\columnwidth]{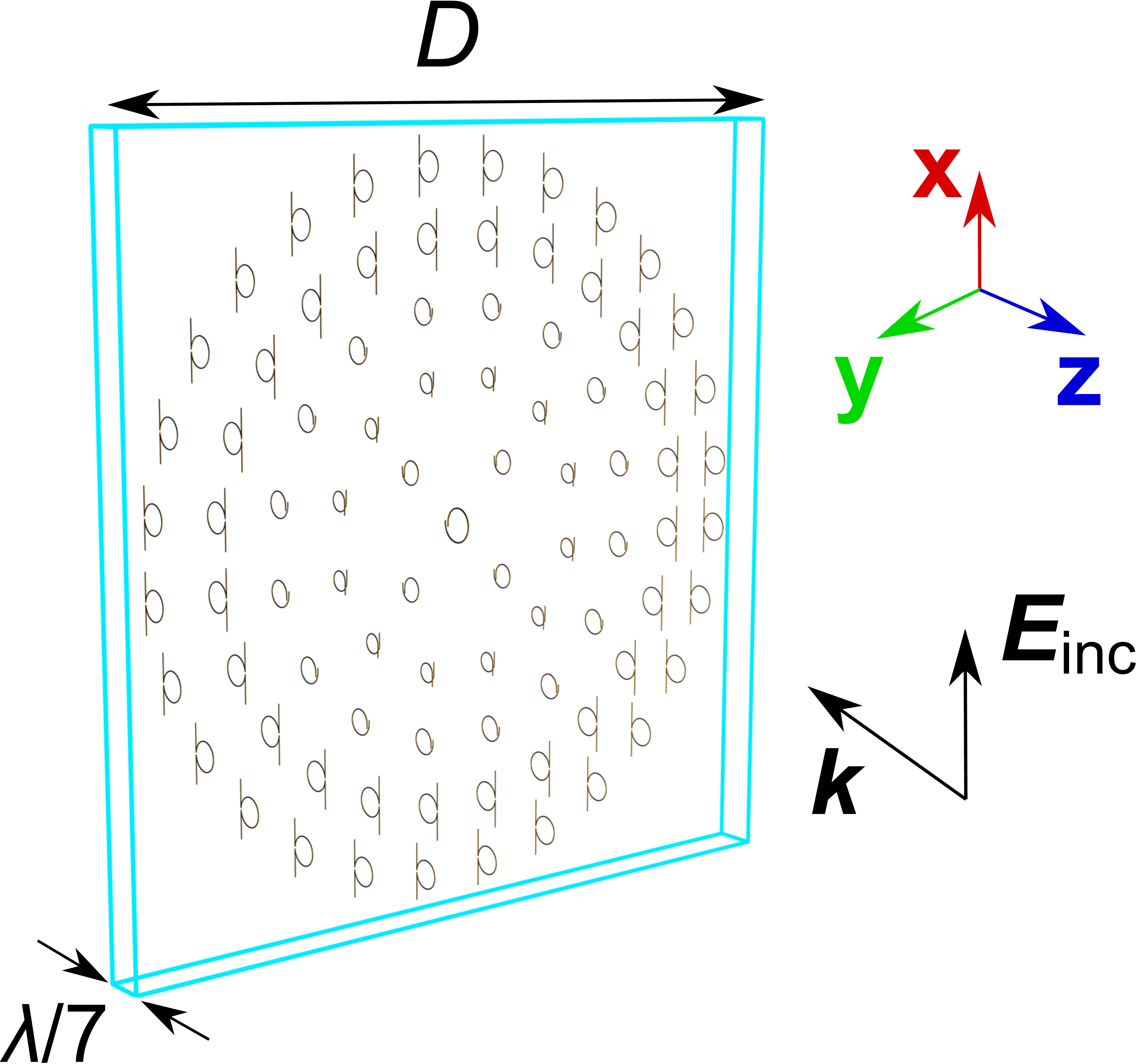}
  \caption{}
  \label{fig:fig3a}
\end{subfigure}\hspace{2mm}
\begin{subfigure}{0.31\columnwidth}
  \centering
  \includegraphics[width=\columnwidth]{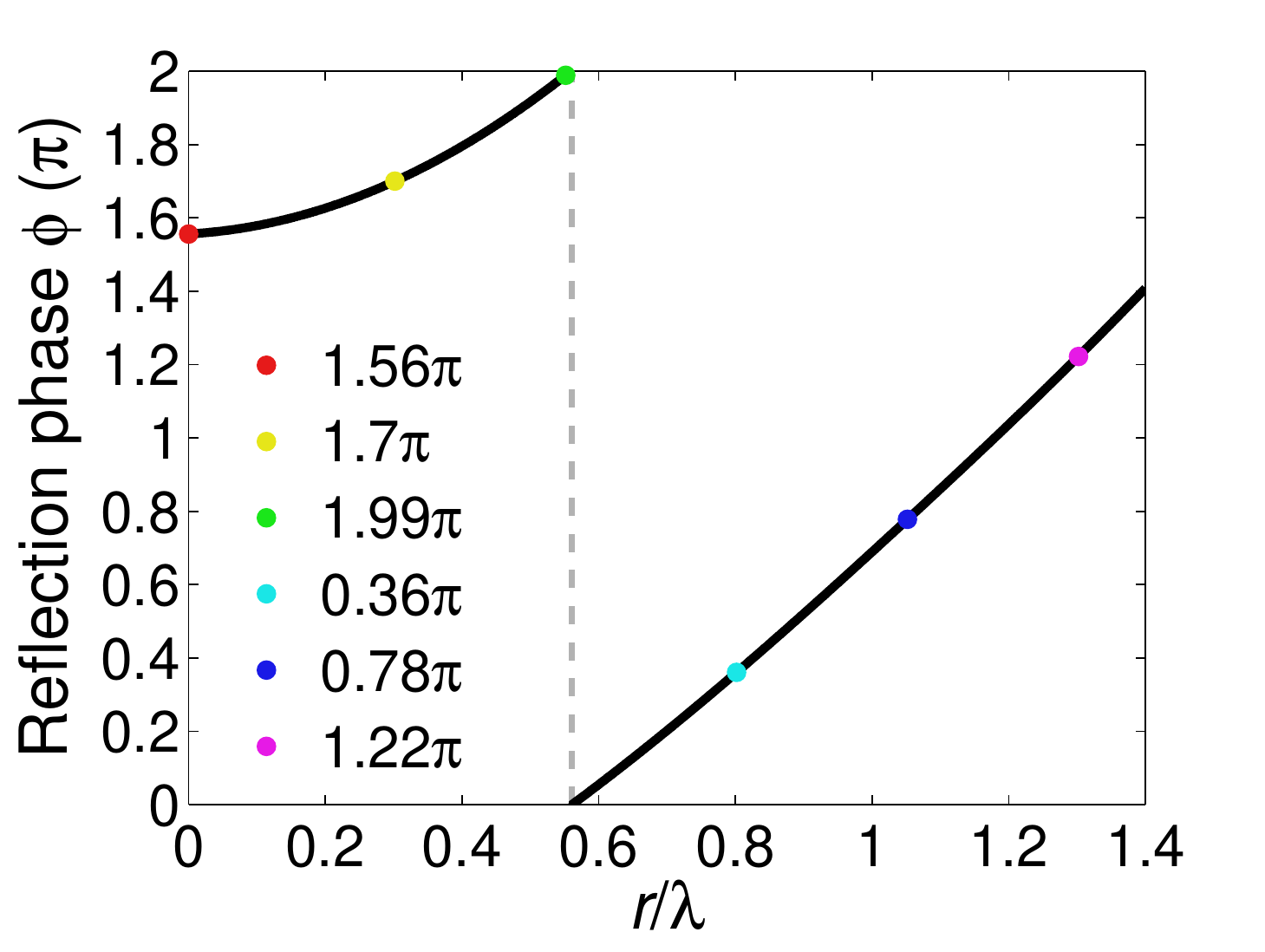}
  \caption{}
  \label{fig:fig3b}
\end{subfigure}\hspace{2mm}
\begin{subfigure}{0.31\columnwidth}
  \centering
  \includegraphics[width=\columnwidth]{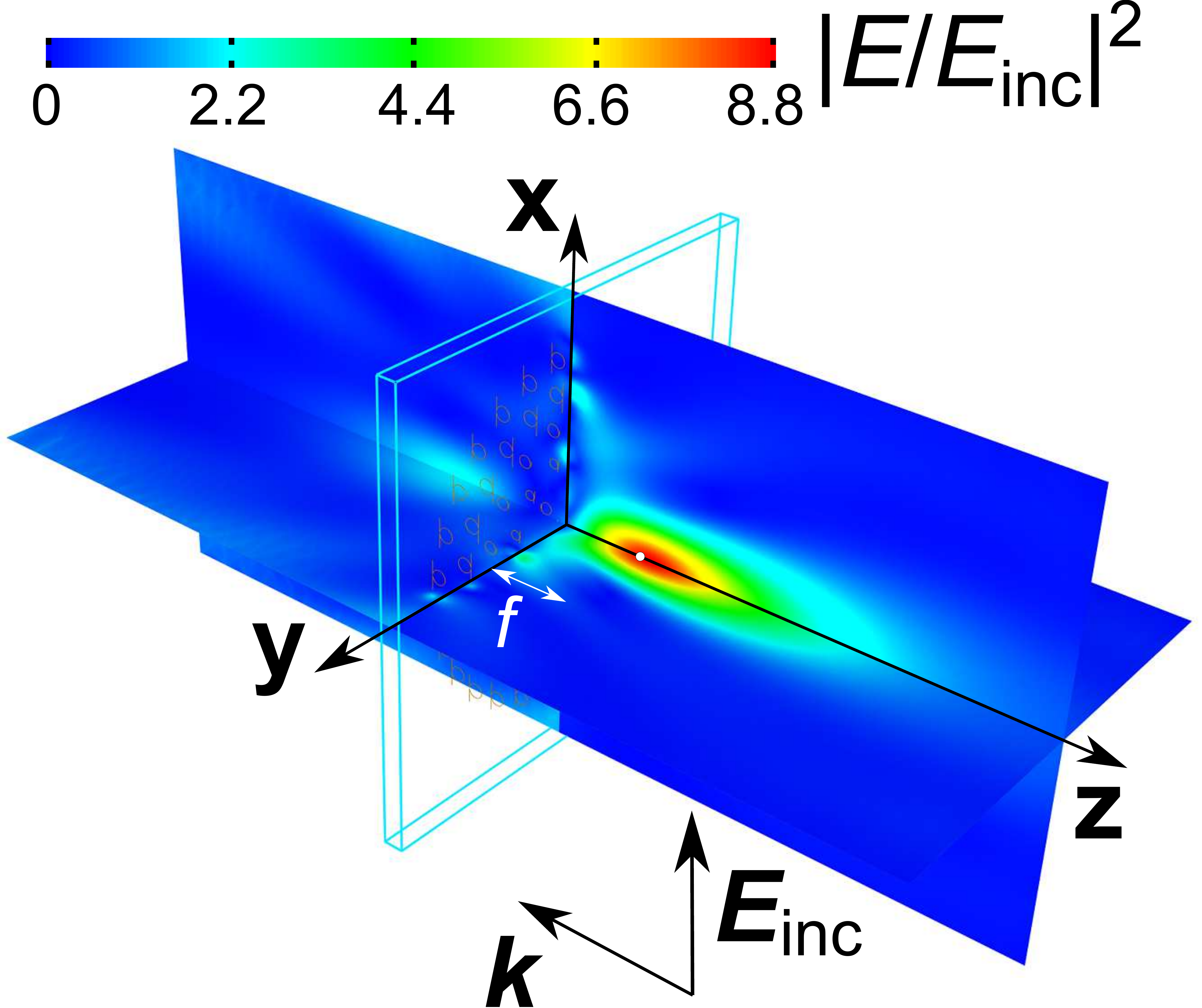}
  \caption{}
  \label{fig:fig3c}
\end{subfigure}%
\caption{Focusing at sub-wavelength distances. (a) A metalens composed of 6 concentric arrays of the designed sub-wavelength inclusions. The dielectric substrate is denoted by the blue box. (b) The solid curve shows the required phase variations at the metalens surface versus the distance to the center of the lens $r$. The dots of different colors show the phase that is provided by the designed inclusions in the different concentric arrays. (c) Power density distribution (normalized to the incident power density) of the transmitted (the $-z$~half-space) and reflected (the $+z$~half-space) waves. The power intensity maps are depicted on two orthogonal cross-section planes (the $xz$ and $yz$-planes).}
\label{fig:fig3}
\end{figure*}
The diameter of the lens is $D=2.8\lambda$ and the effective thickness of the structure (thickness of the inclusions) is $t=\lambda/7$.
The mechanically supporting dielectric is a Styrofoam plate ($\epsilon_{\rm r}=1.03$, $\rm tan \delta=0.0001$) of 10~mm thickness.
Realization of the desired lens response requires certain reflection phase variations
along the surface ensuring the scattered fields from all the
inclusions constructively interfere at the desired point. Without loss of generality, we content ourselves with the case when the metalens confines energy on its axial line at a focal distance $f$. The required phase of reflection $\phi$ for an inclusion located at a distance $r$ from the center is
\begin{equation}
\displaystyle
\phi(r)=\phi_0+\frac{\omega}{c}\sqrt{r^2+f^2},
\label{eq:2}
\end{equation}
\noindent
where $c$ is the speed of light, $\phi_0$ is an additional constant phase that can be chosen arbitrarily. Equation (\ref{eq:2}) shows that a metalens with a smaller focal distance requires faster phase variations along the surface. In order to have enough smooth phase gradient to allow surface homogenization, we choose the focal length of $f=0.6\lambda$ for a metalens consisting of wire inclusions of $\lambda/7$ size. The relation between  phase $\phi$ and coordinate $r$ of the inclusions of a metalens operating at 5~GHz is plotted in Fig.~\ref{fig:fig3b}. 
For the design convenience we have chosen the additional phase to be $\phi_0=\pi/3$.

An incident plane wave impinges on the metalens from the $+z$-direction with the electric field along the $x$-axis. 
Figure~\ref{fig:fig3c} shows the simulated power distribution of the reflection and transmission from the metalens. Very weak but non-zero transmission in the far-zone is caused mainly by diffraction effects on the edges of the metasurface. The metalens effectively reflects the wave and focuses it tightly near
the diffraction limit to a spot of only $2.8\lambda\times0.9\lambda$
size ($1/e^2$ beamwidth). The extremely strong focusing ability of
the designed metalens provides the focal length of only $f=0.65\lambda$ and high energy gain of 8.8 in the spot. The unprecedentedly short focal length of our metalens is significantly smaller than those of conventional and other metamaterial-based lenses\cite{alu,babinet,grin}. Similar strong focusing abilities have been recently achieved with a lens based on hetero-junctions of anisotropic metamaterials\cite{hetero}. However, that construction is optically thick and generally operates only with incident TM-polarized light, which reduces the efficiency of the lens. 
The f-number (the ratio of the focal length to the aperture diameter of a lens $f/D$) for the designed metalens comes to 0.23 and can be further decreased by increasing the aperture, which is in this example only $2.8\lambda$. Such small f-number of the metalens allows of gathering more power and generally provides a brighter image. 
 It should be noted that the metalens possesses asymmetrical focusing properties with respect to the propagation direction of incident waves.  The performance of the lens for incident waves from the opposite direction ($-z$-direction) is depicted in Supplementary Fig.~S4.

The proposed metalens can find various applications within the entire electromagnetic spectrum due to its unique characteristics: for example, replacing parabolic dish antennas with planar and low-profile structures, micrometer-scale light concentrators, and integrated optical lenses.
Another feature of the metalens to become practically transparent at frequencies outside of the operating frequency band can be advantageous for specific applications.

\subsection{Experimental realization and characterization.}
The operation of the two proposed metamirrors was verified by conducting measurements inside a parallel-plate waveguide\cite{pekka}. The sub-wavelength inclusions of the metamirrors can be represented as coupled vertical electric and horizontal magnetic dipoles. According to the image theory, by placing a one-dimensional array of such inclusions inside a parallel-plate waveguide with the height equal to the array period, we can effectively emulate the two-dimensional scenario in free space where the structure is infinitely periodic in the vertical $x$-direction. 
The incident waves with the $x$-oriented electric field are generated by a vertical coaxial feed, as described in more detail in Methods. At the operating frequency of the metamirrors only TEM waves (with electric and magnetic fields orthogonal to the propagation direction) can propagate without attenuation in the waveguide. 
Spatial distribution of the $x$-component (the other components are negligible with the described incidence and geometry) of the electric field inside the waveguide is measured through a copper mesh embedded in the bottom plate using a movable coaxial probe positioned under the mesh. The mesh does not significantly disturb the field distribution inside the waveguide since the mesh period is much smaller than the wavelength.
In order to visualize the reflected (back-scattered) fields from a metamirror, we have conducted two sets of measurements. In the first set, we measure the field distribution of an incident wave propagating in the empty waveguide. The second set implies measurement of the total electric field (incident and reflected) in the waveguide in the presence of the metamirror. Consequently, the reflected field distribution can be achieved by subtracting the incident fields in the first measurement from the total fields in the second measurement.

\begin{figure*}[h]
\centering
\begin{subfigure}{0.48\columnwidth}
  \centering
  \includegraphics[width=\columnwidth]{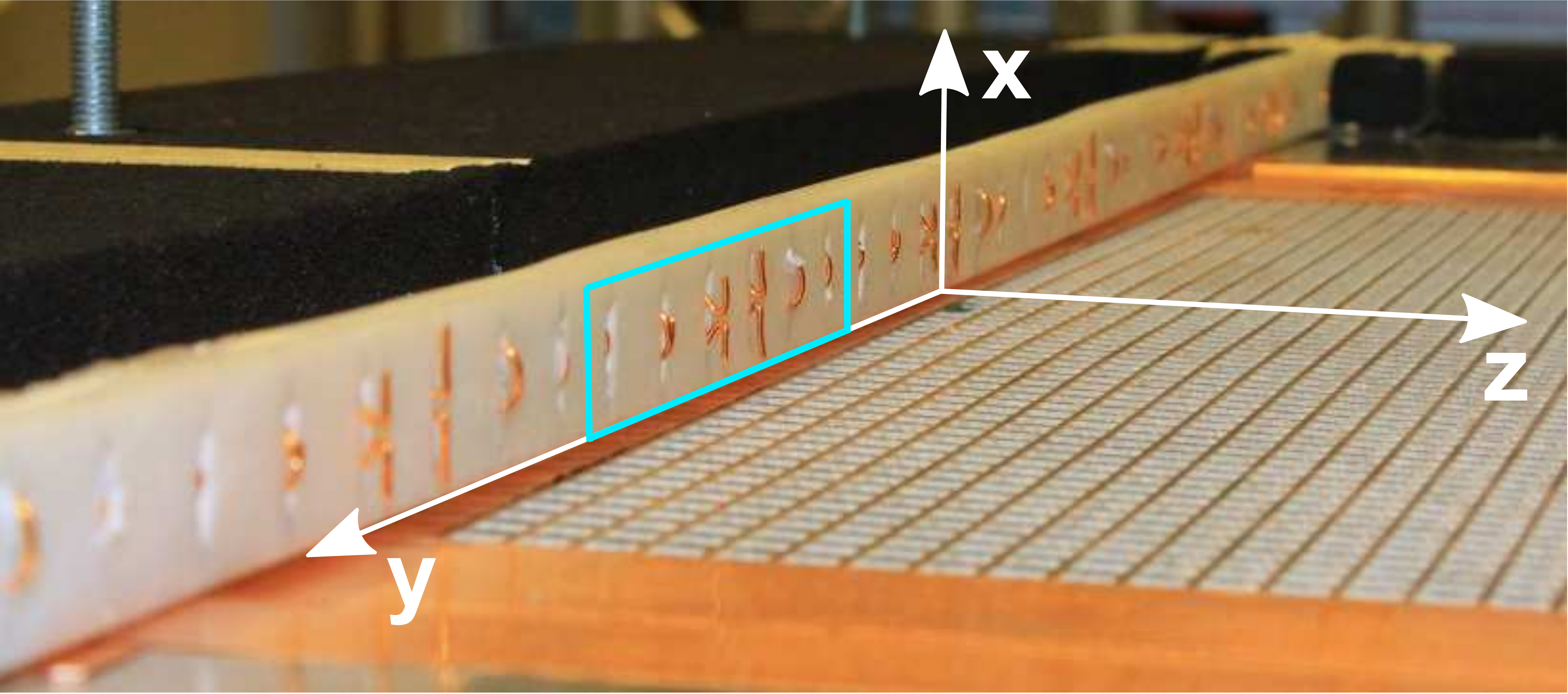}
  \caption{}
  \label{fig:fig4a}
\end{subfigure}\hspace{2mm}%
\begin{subfigure}{0.48\columnwidth}
  \centering
  \includegraphics[width=\columnwidth]{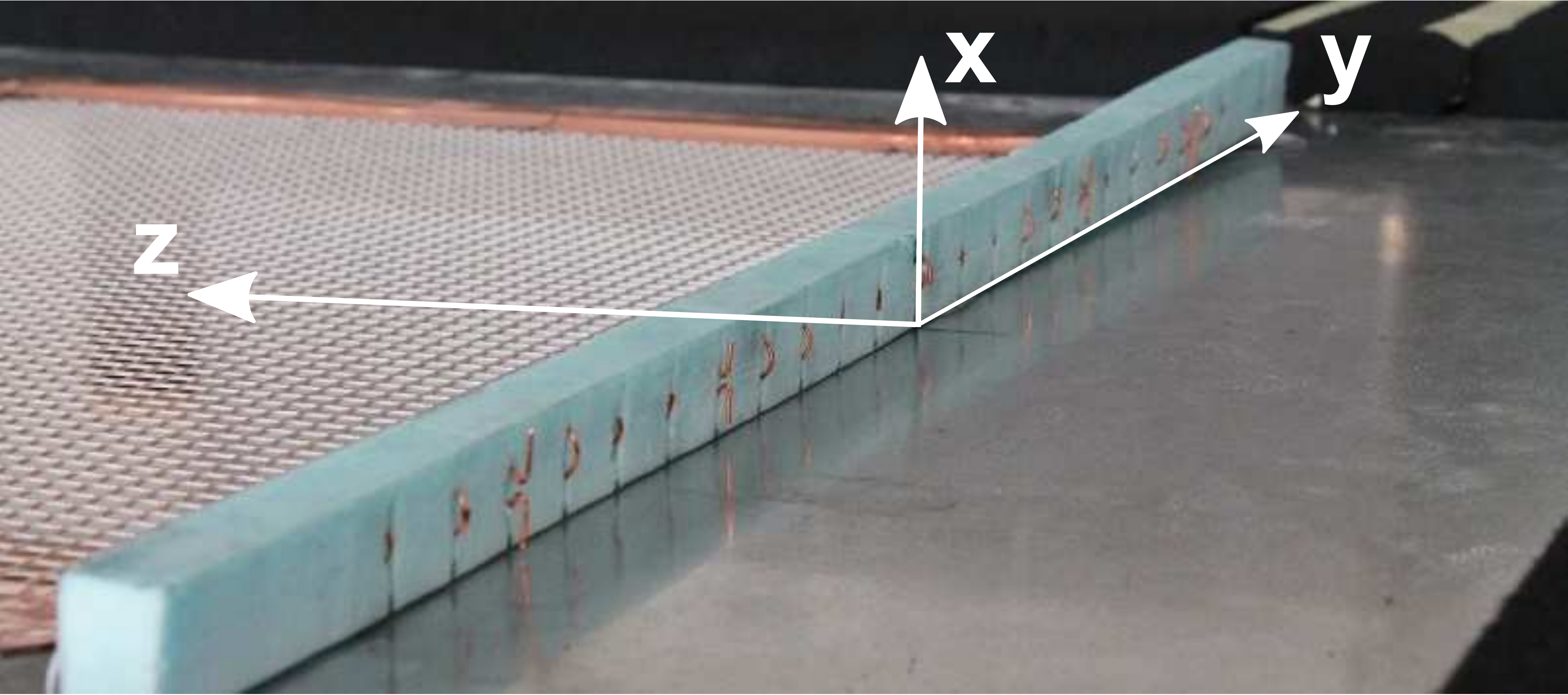}
  \caption{}
  \label{fig:fig4b}
\end{subfigure}
\begin{subfigure}{0.48\columnwidth}
  \centering
  \includegraphics[width=\columnwidth]{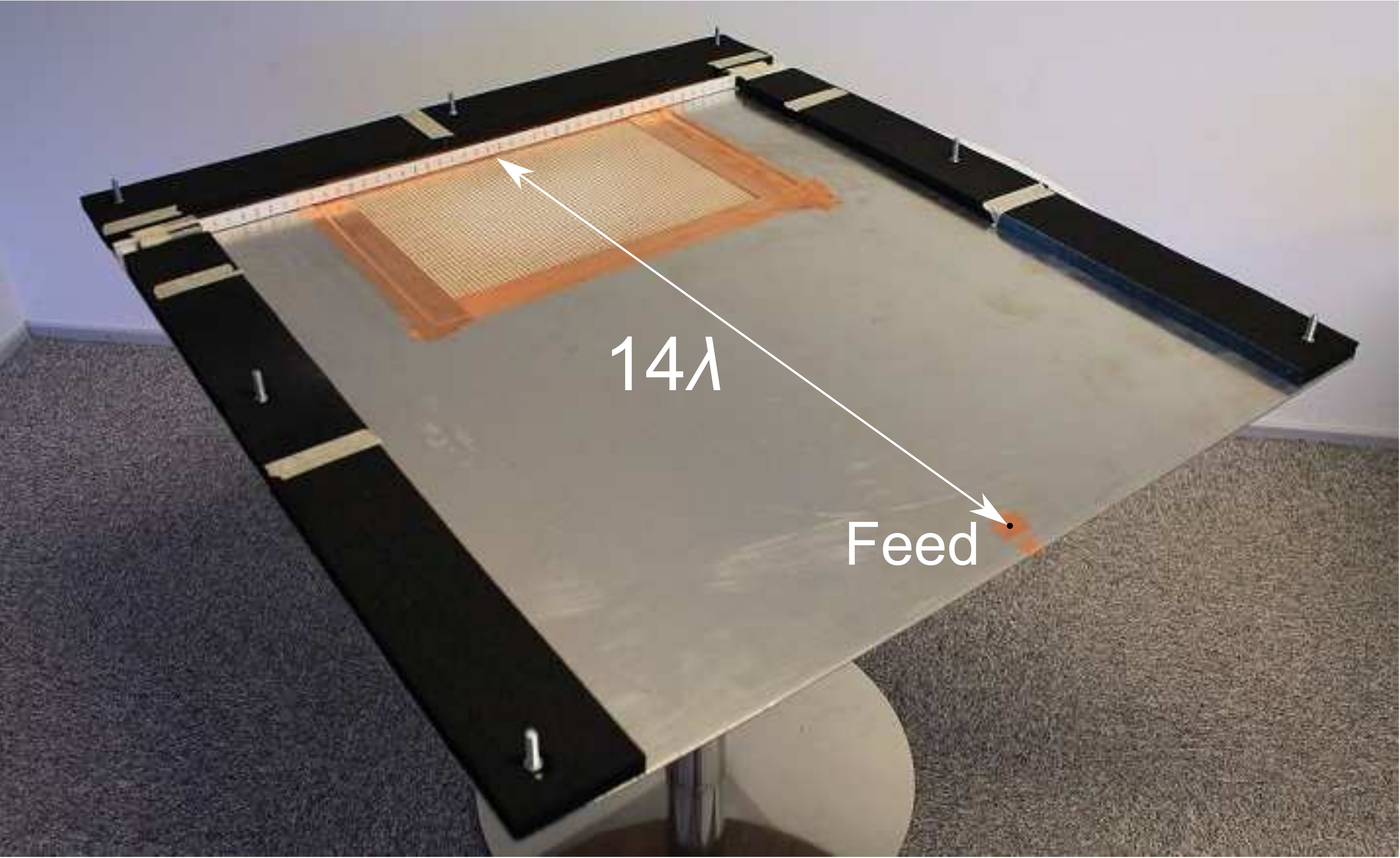}
  \caption{}
  \label{fig:fig4c}
\end{subfigure}\hspace{2mm}%
\begin{subfigure}{0.48\columnwidth}
  \centering
  \includegraphics[width=\columnwidth]{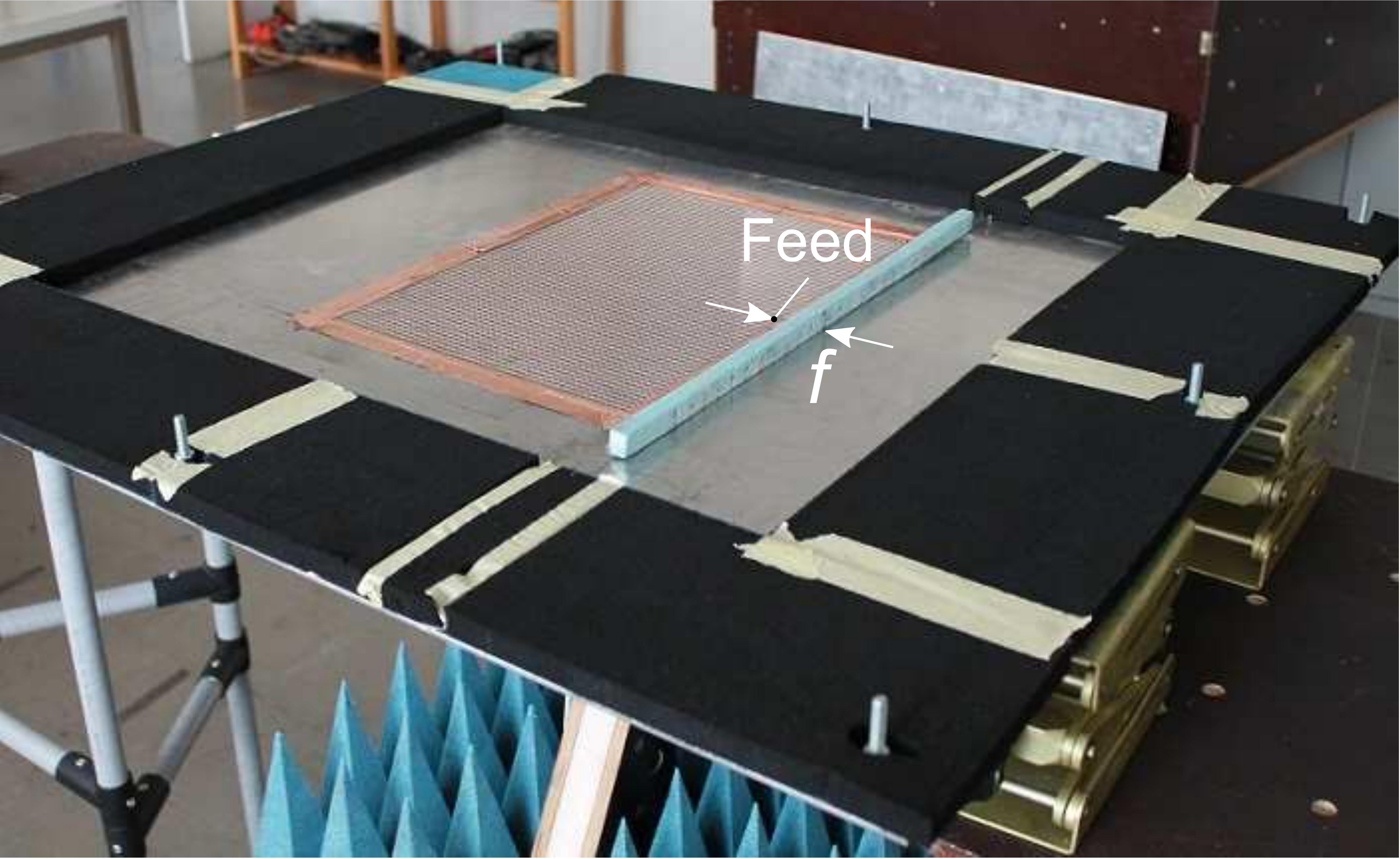}
  \caption{}
  \label{fig:fig4d}
\end{subfigure}
\caption{Experimental realization. (a) Fabricated metamirror for changing direction of the reflection wavefronts. The blue box shows the period of the metamirror along the $x$ and $y$-axis. (b) Fabricated metalens consisted of 23 sub-wavelength inclusions. (c-d) Experimental setups for metamirrors in (a) and (b), respectively. The top metal plates of the waveguides are removed.}
\label{fig:fig4}
\end{figure*}

\begin{figure*}[h]
\centering
\begin{subfigure}{0.33\columnwidth}
  \centering
  \includegraphics[width=\columnwidth]{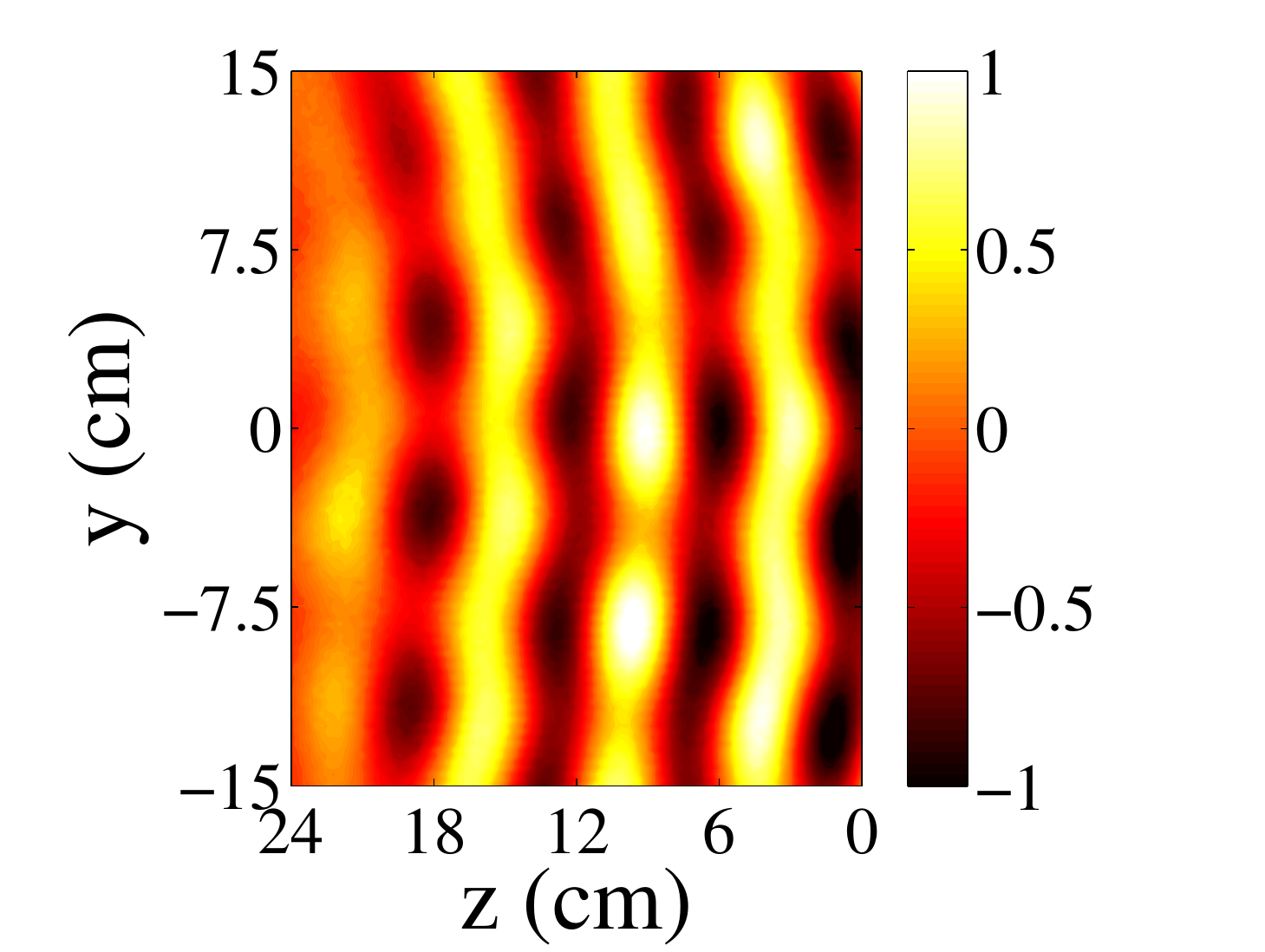}
  \caption{}
  \label{fig:fig5a}
\end{subfigure}%
\begin{subfigure}{0.33\columnwidth}
  \centering
  \includegraphics[width=\columnwidth]{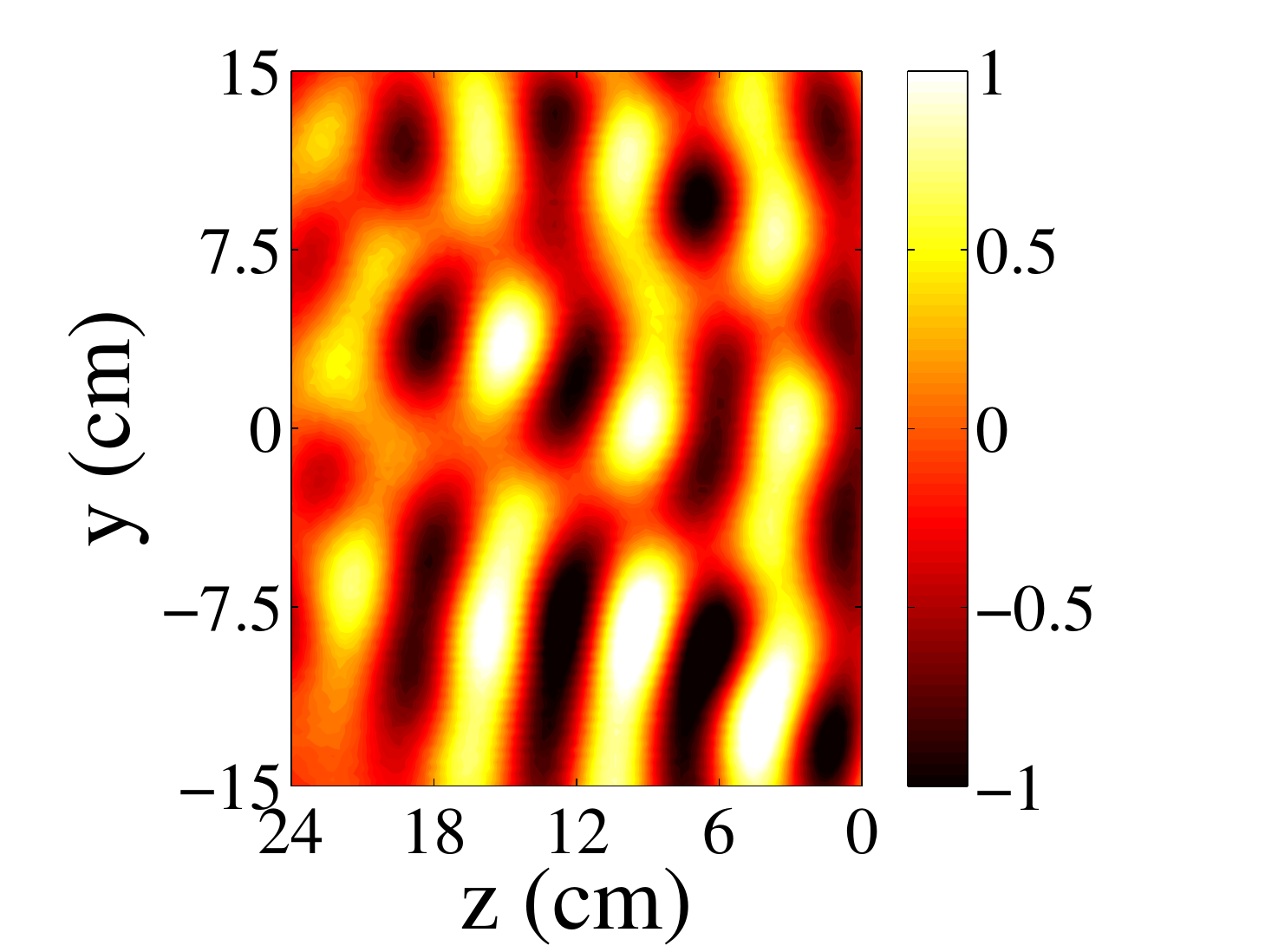}
  \caption{}
  \label{fig:fig5b}
\end{subfigure}%
\begin{subfigure}{0.33\columnwidth}
  \centering
  \includegraphics[width=\columnwidth]{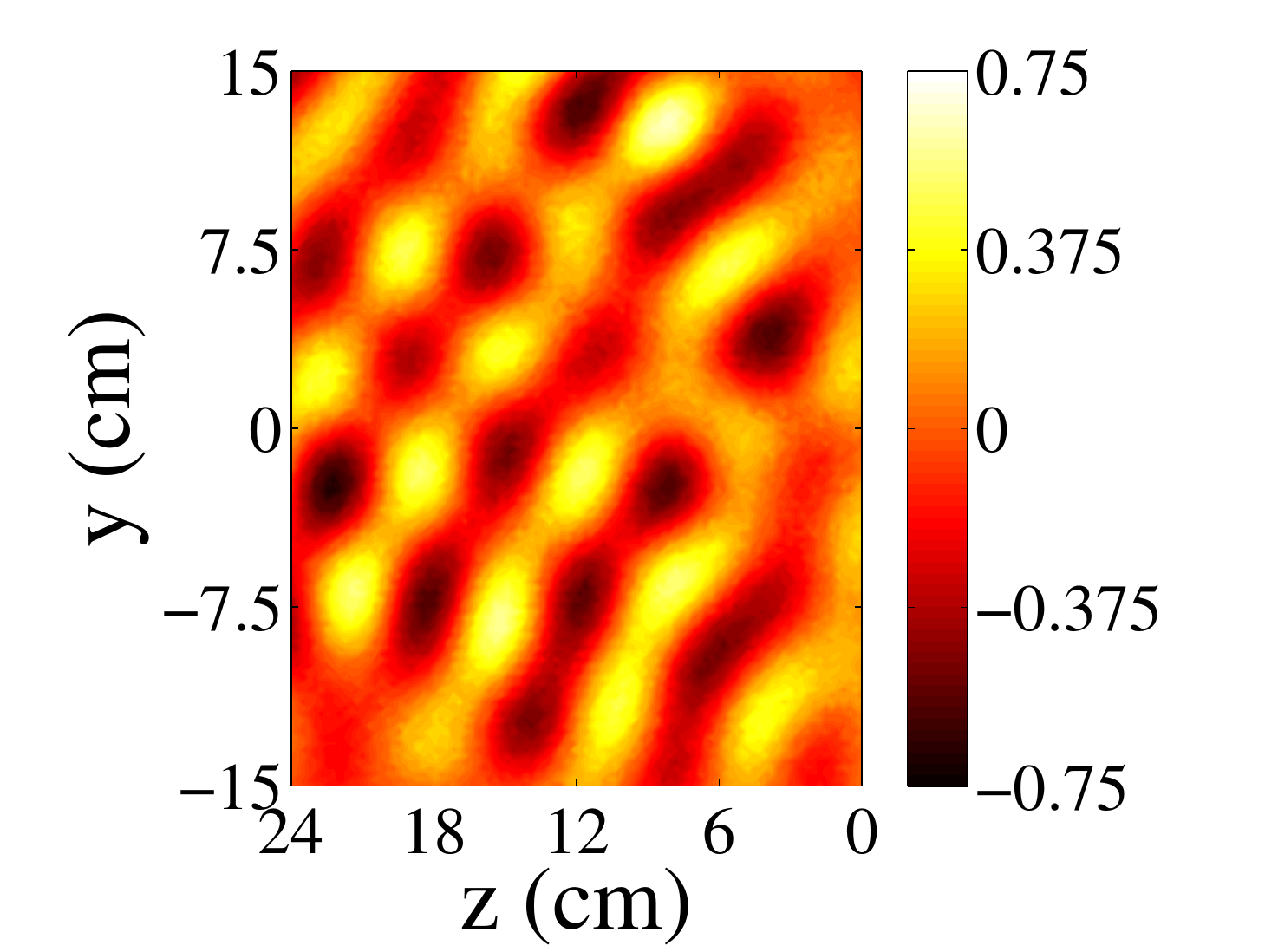}
  \caption{}
  \label{fig:fig5c}
\end{subfigure}
\begin{subfigure}{0.33\columnwidth}
  \centering
  \includegraphics[width=\columnwidth]{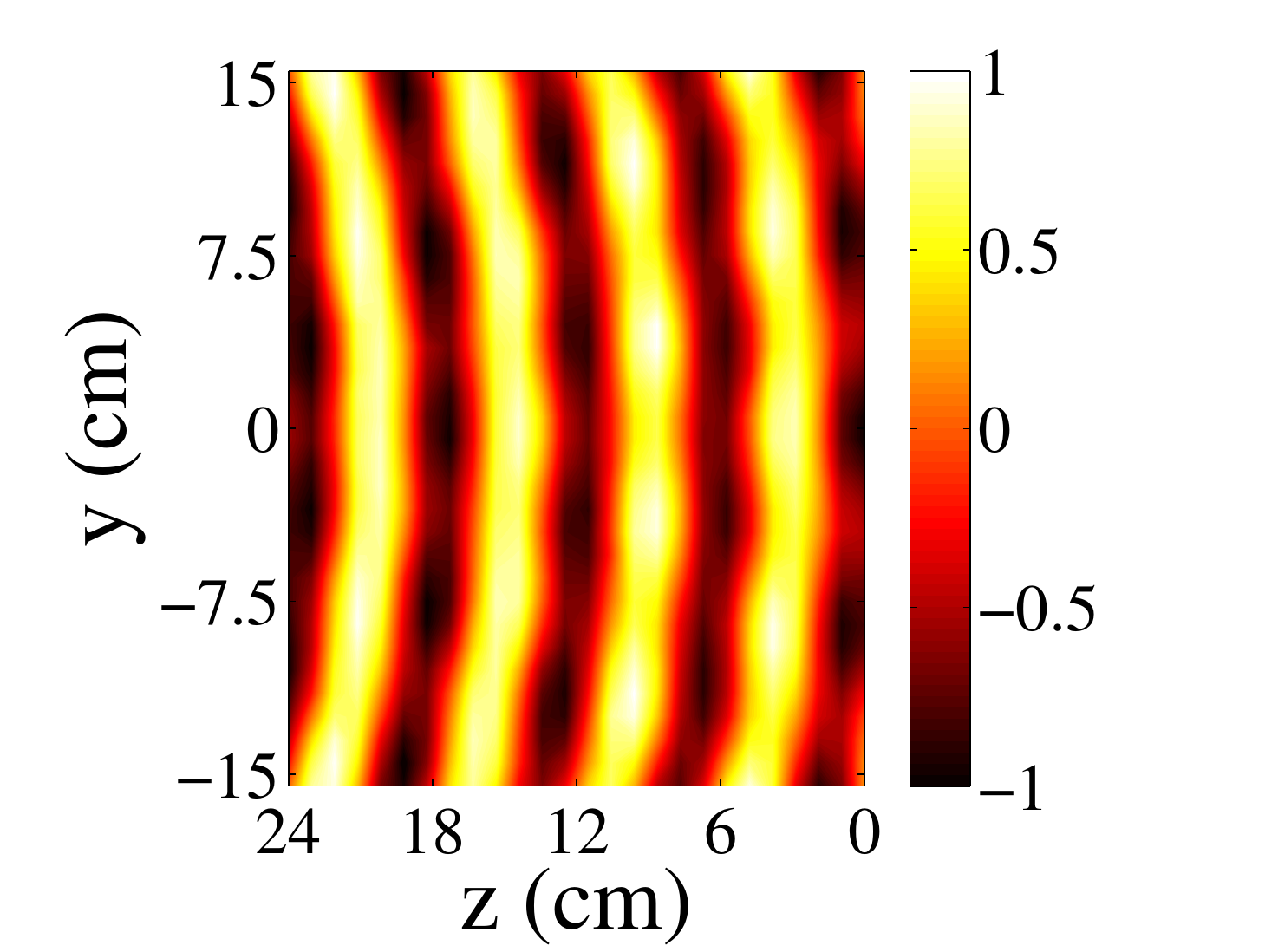}
  \caption{}
  \label{fig:fig5d}
\end{subfigure}%
\begin{subfigure}{0.33\columnwidth}
  \centering
  \includegraphics[width=\columnwidth]{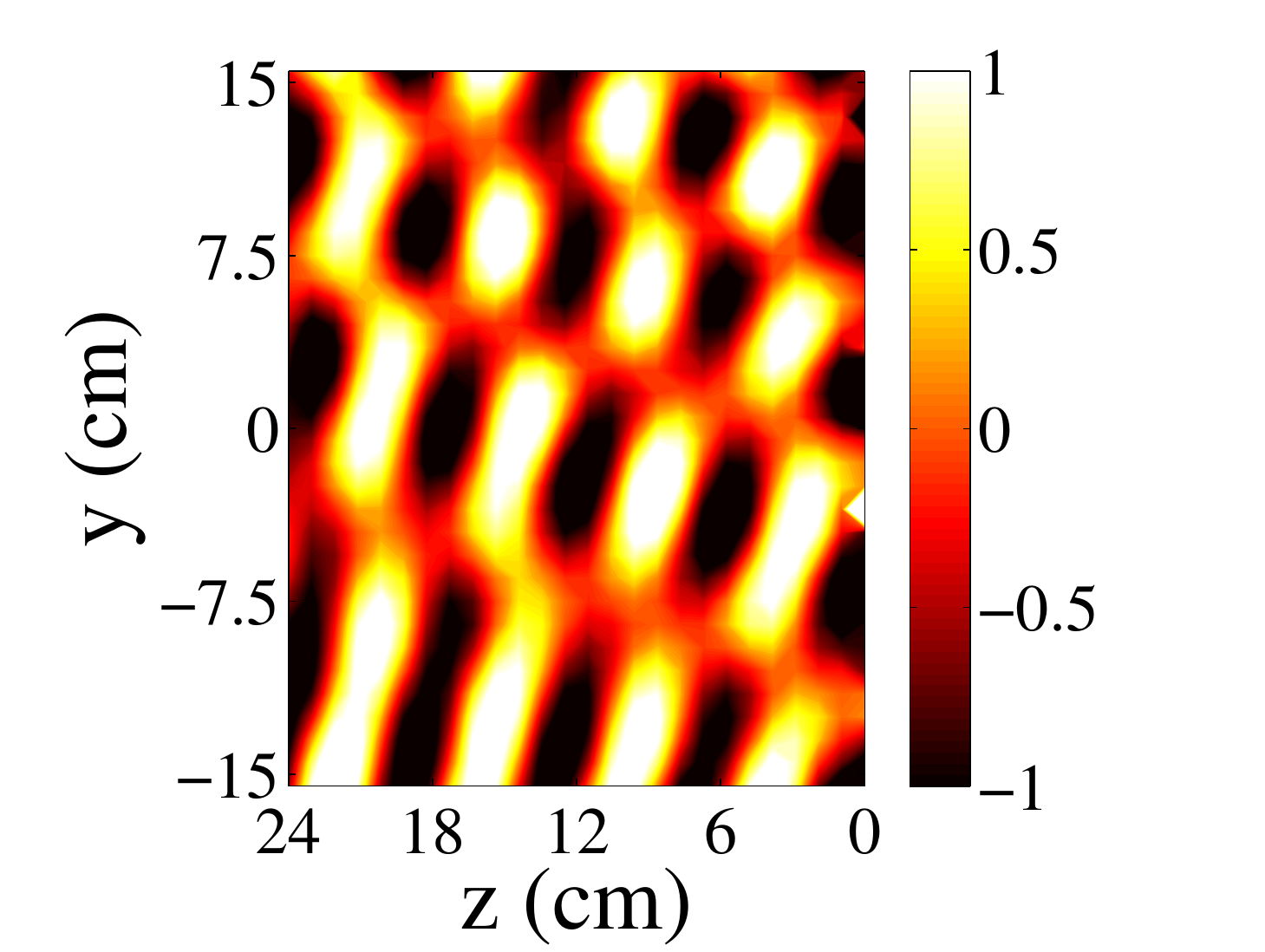}
  \caption{}
  \label{fig:fig5e}
\end{subfigure}%
\begin{subfigure}{0.33\columnwidth}
  \centering
  \includegraphics[width=\columnwidth]{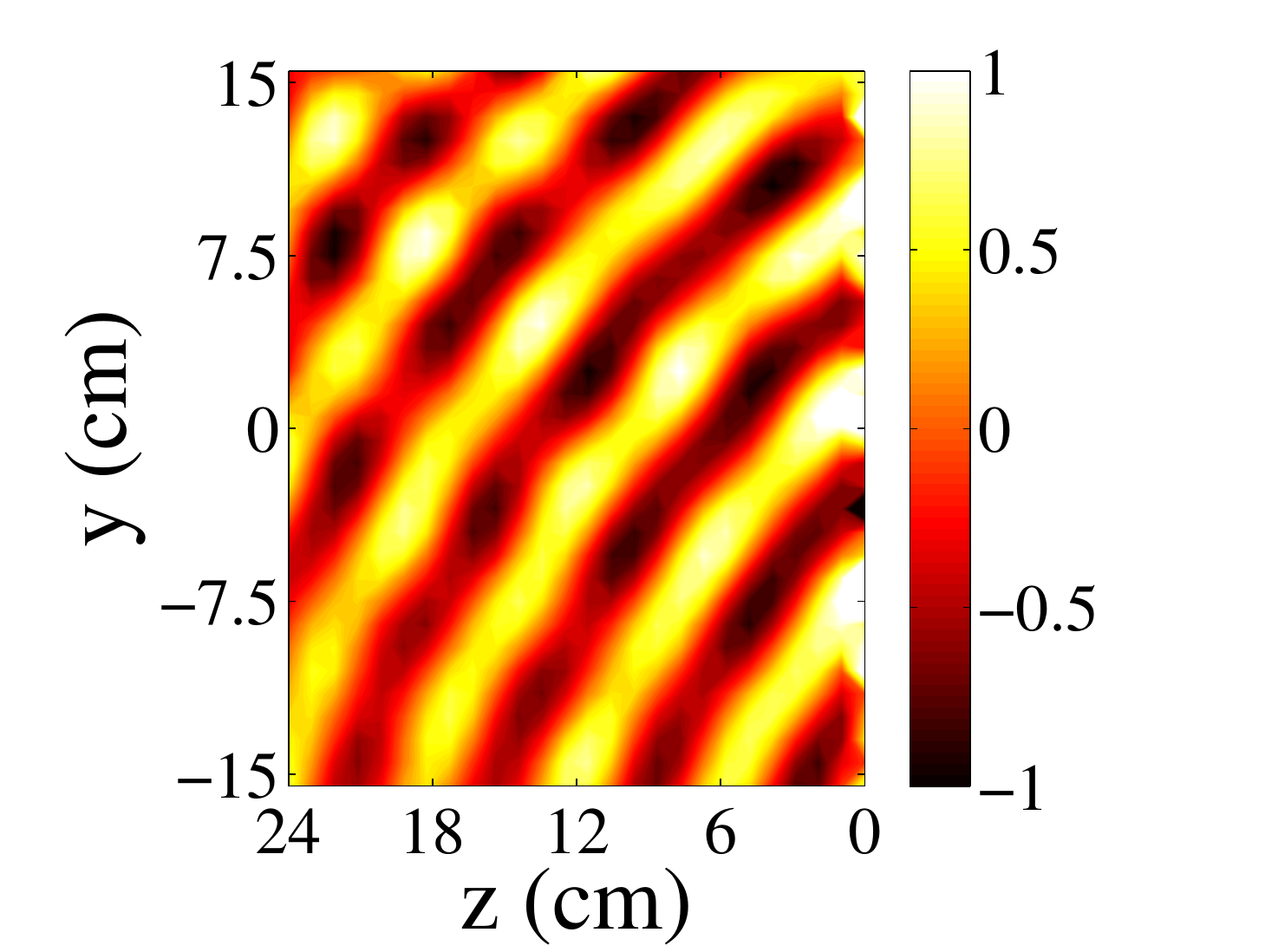}
  \caption{}
  \label{fig:fig5f}
\end{subfigure}%
\caption{Experimental verification of the anomalous reflection. (a-c) Distribution of the measured co-polarized electric field in the waveguide for a metamirror reflecting normally incident waves at an angle $\theta=45^{\circ}$. (a) The field distribution of an incident wave in the empty waveguide. (b) The field distribution in the presence of the metamirror (interference of the incident and reflected waves). (c) The field distribution of the reflected wave. (e-f) Corresponding electric field distributions from full-wave numerical simulations of the measurements in a waveguide.}
\label{fig:fig5}
\end{figure*}

First, we analyze  the metamirror reflecting normally incident plane waves to an angle $\theta=45^{\circ}$ from the normal. The characteristics of the metamirror are specified in Methods. The manufactured metamirror (Fig.~\ref{fig:fig4a}) is positioned at a distance of 80~cm (about 14~operating wavelengths) from the feed (Fig.~\ref{fig:fig4c}). 
At this distance, cylindrical waves radiated by the feed can be considered as nearly plane waves.
The electric field distribution of the incident, total and extracted reflected waves from the metamirror at a frequency 5.2~GHz are shown in Fig.~\ref{fig:fig5a}--\ref{fig:fig5c}. 
Figures~\ref{fig:fig5d}--\ref{fig:fig5f} show the corresponding field distributions achieved with full-wave numerical simulations of the experimental setup at the  frequency 5~GHz. The metamirror totally reflects a normally incident plane wave from the $+z$-direction (Figs.~\ref{fig:fig5a} and \ref{fig:fig5d}) to an angle $\theta=45^{\circ}$ from the normal (Figs.~\ref{fig:fig5c} and \ref{fig:fig5f}).
The measured and simulated results are in good agreement and manifest the desired functionality of the metamirror predicted by our theory. The reduced amplitude of the reflected wavefront, as seen from Fig.~\ref{fig:fig5e}, can be caused by insufficient sensitivity of the measuring probe and the presence of non-zero horizontal components of the total electric field in the waveguide. The small difference of the optimal operating frequencies for the measured and simulated data can be explained by inaccuracies in the dimensions of the manufactured inclusions.

\begin{figure*}[h]
\centering
\begin{subfigure}{0.33\columnwidth}
  \centering
  \includegraphics[width=\columnwidth]{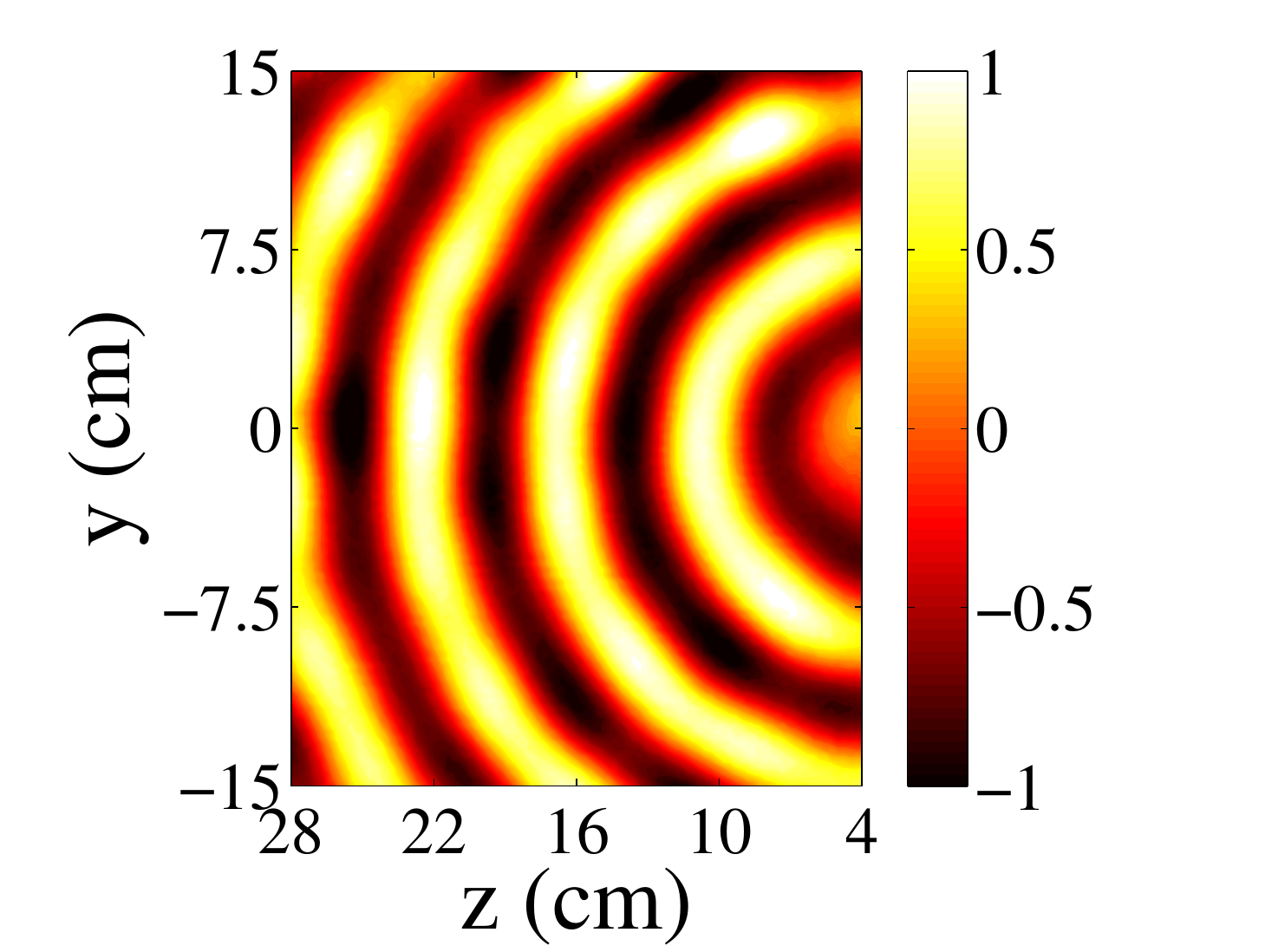}
  \caption{}
  \label{fig:fig6a}
\end{subfigure}%
\begin{subfigure}{0.33\columnwidth}
  \centering
  \includegraphics[width=\columnwidth]{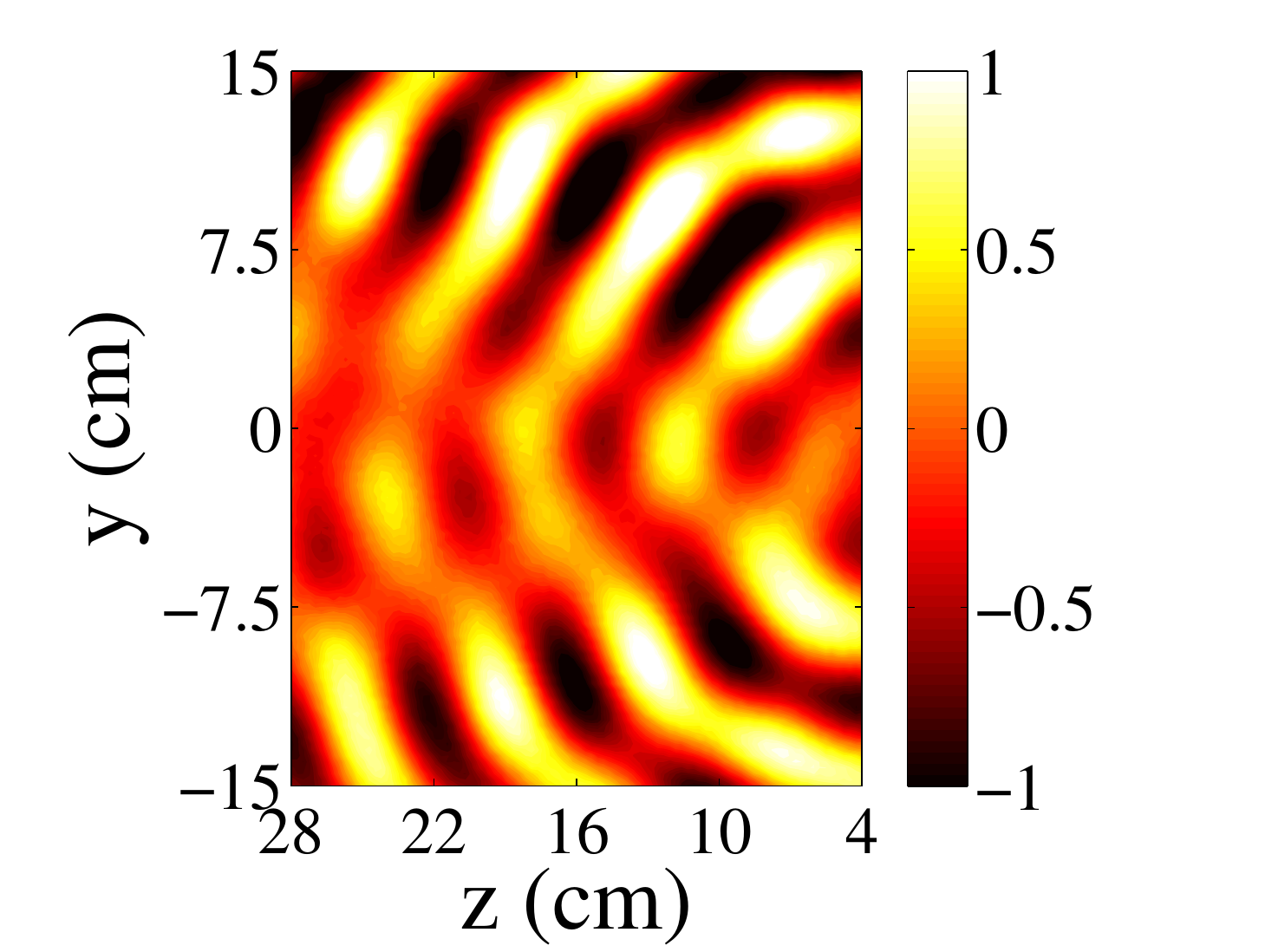}
  \caption{}
  \label{fig:fig6b}
\end{subfigure}%
\begin{subfigure}{0.33\columnwidth}
  \centering
  \includegraphics[width=\columnwidth]{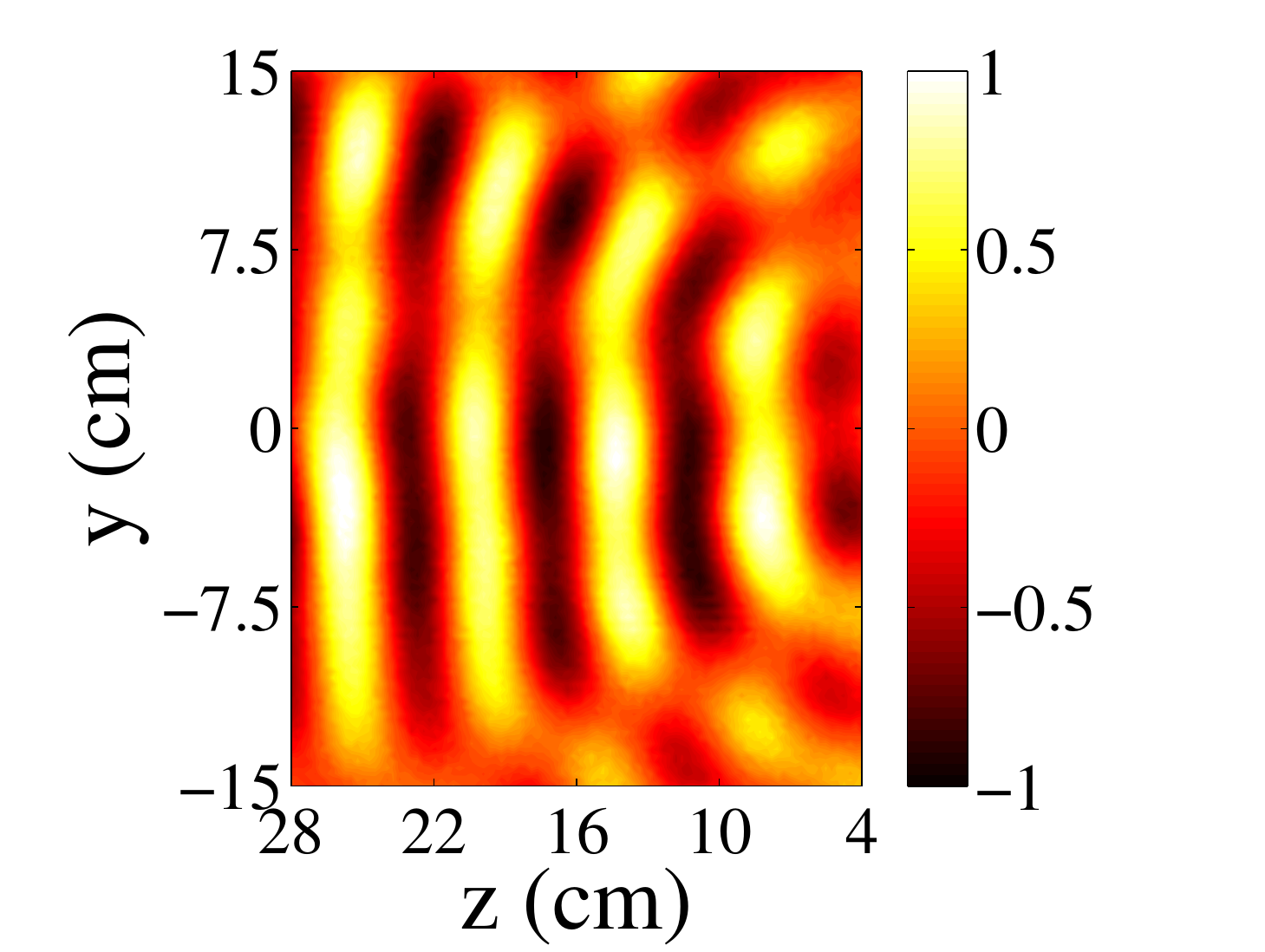}
  \caption{}
  \label{fig:fig6c}
\end{subfigure}
\begin{subfigure}{0.33\columnwidth}
  \centering
  \includegraphics[width=\columnwidth]{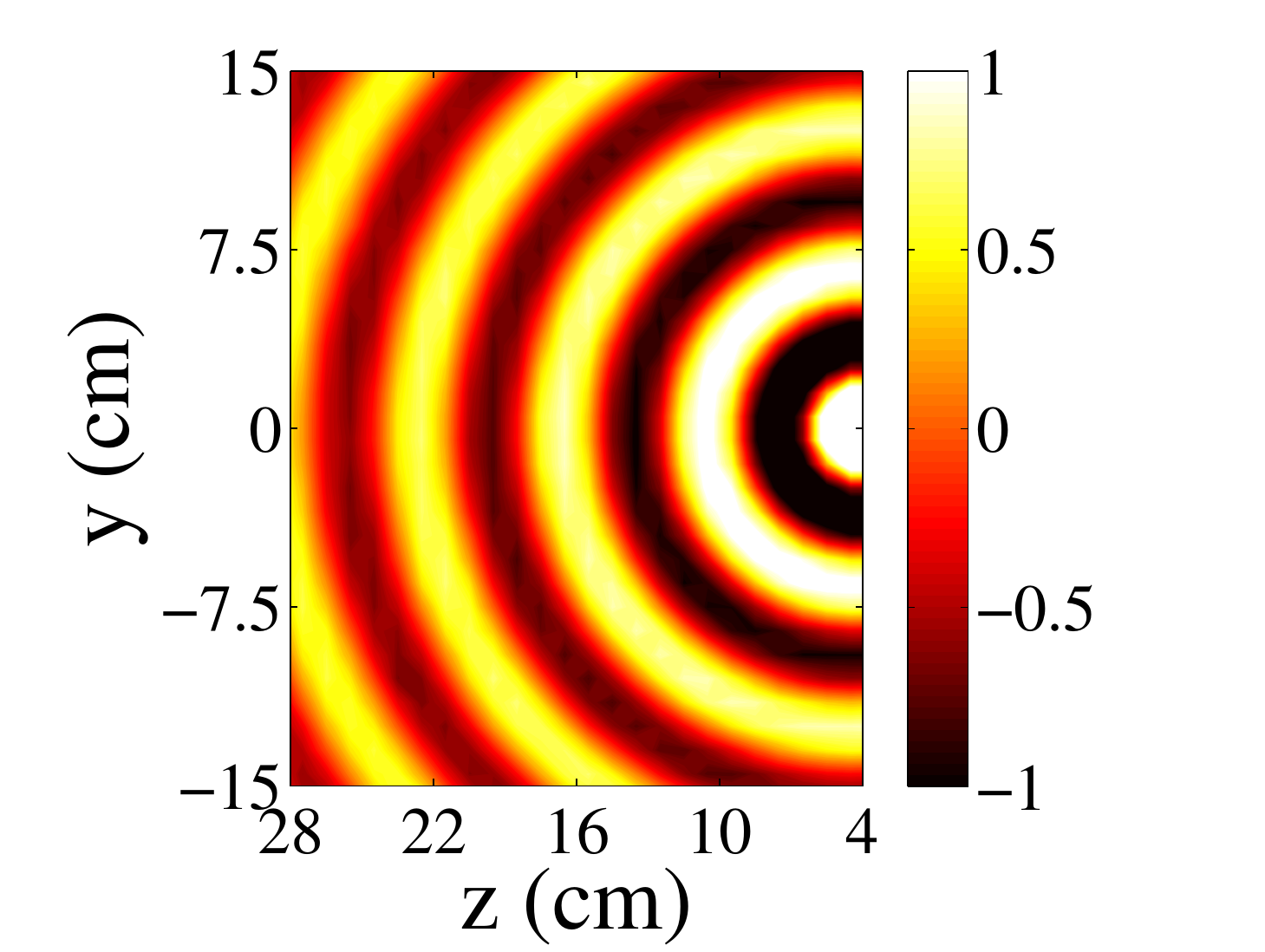}
  \caption{}
  \label{fig:fig6d}
\end{subfigure}%
\begin{subfigure}{0.33\columnwidth}
  \centering
  \includegraphics[width=\columnwidth]{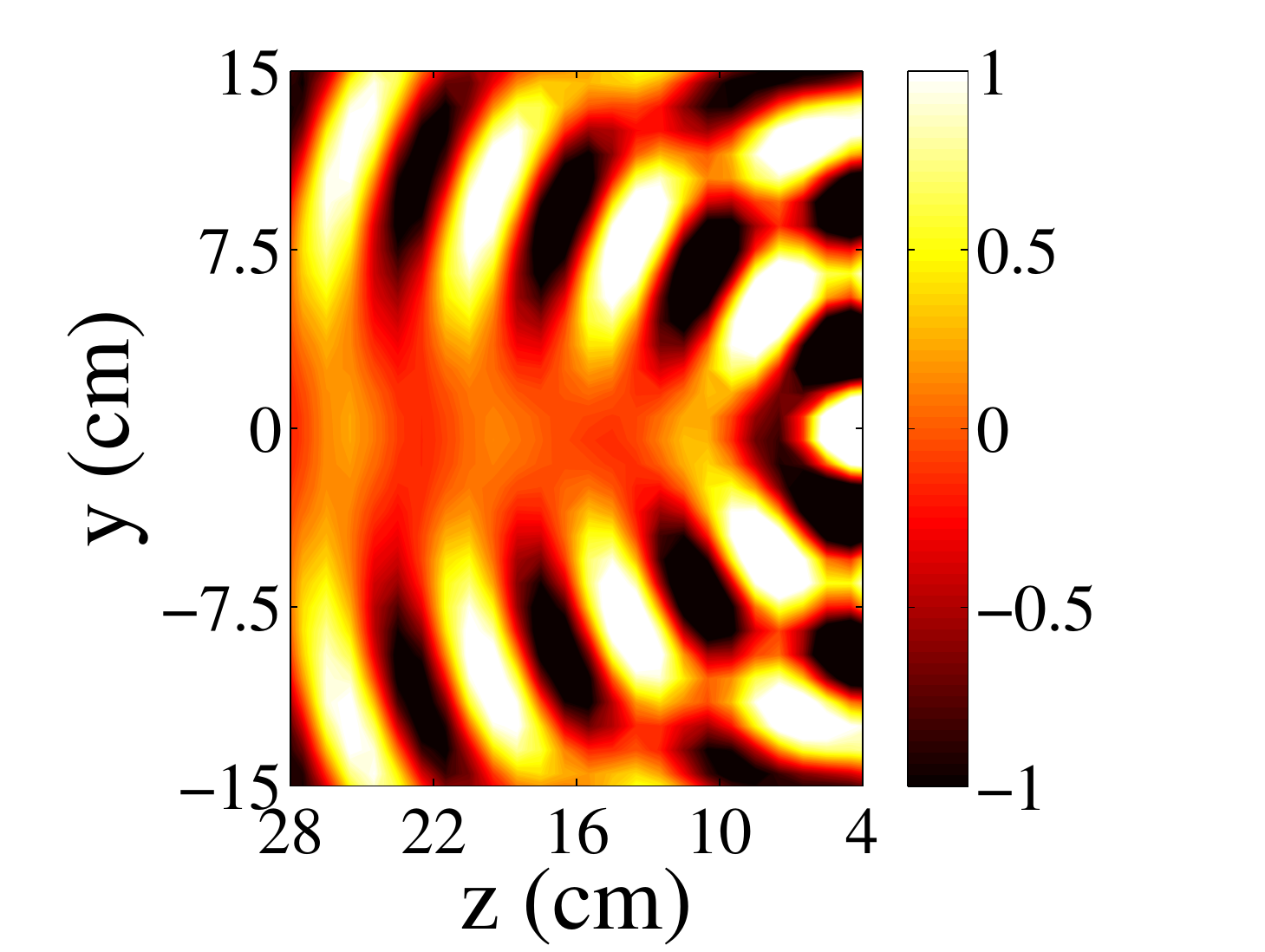}
  \caption{}
  \label{fig:fig6e}
\end{subfigure}%
\begin{subfigure}{0.33\columnwidth}
  \centering
  \includegraphics[width=\columnwidth]{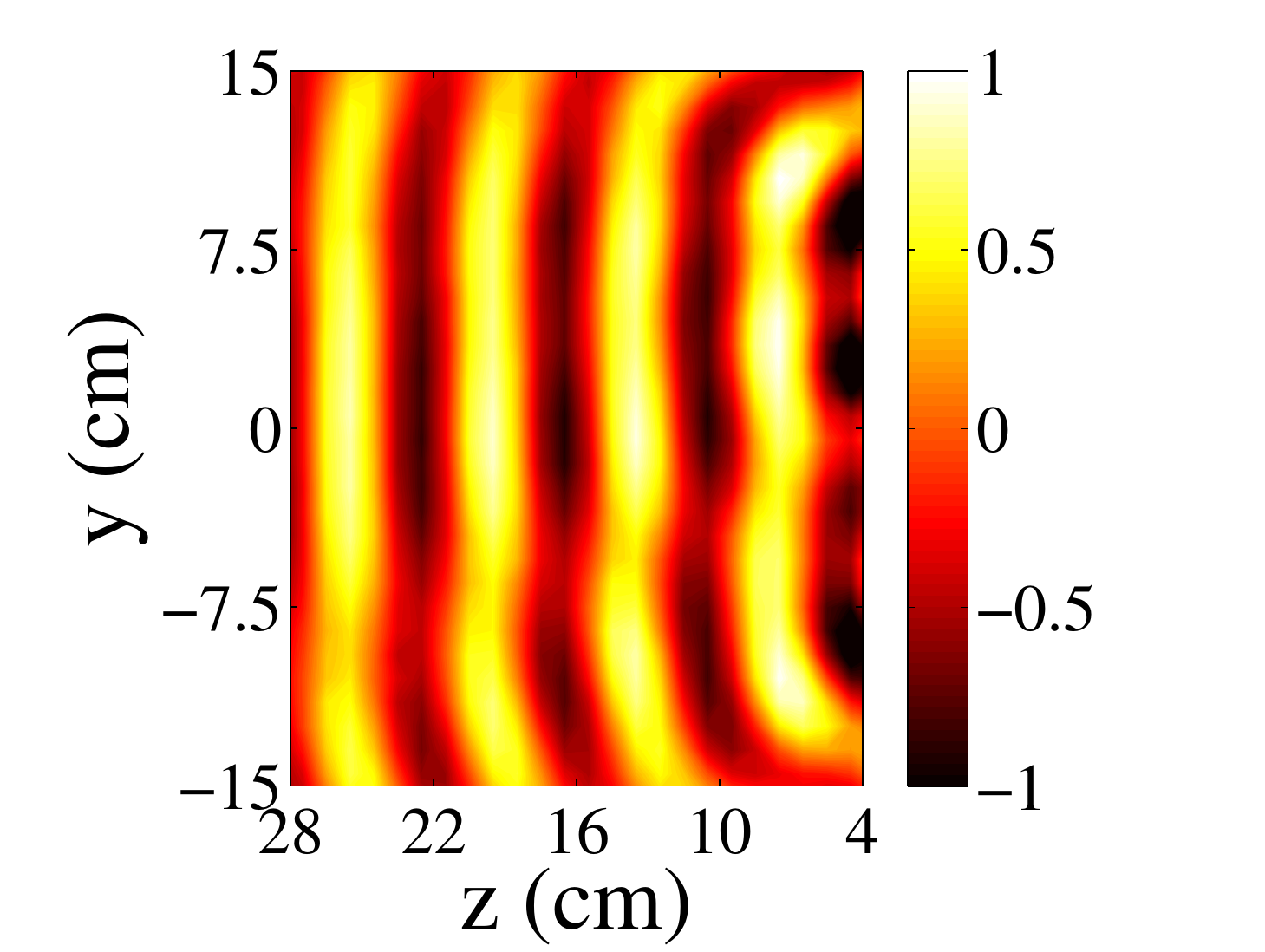}
  \caption{}
  \label{fig:fig6f}
\end{subfigure}%
\caption{Experimental verification of the metalens operation. (a-c) Distribution of the measured co-polarized electric field for a metalens in the waveguide. (a) The field distribution of an incident cylindrical wave radiated by the feed in the empty waveguide. The feed is positioned at the focal point of the lens. (b) The field distribution in the presence of the metalens (interference of the cylindrical incident and plane reflected waves). (c) The field distribution of the reflected wave. (e-f) Corresponding electric field distributions from full-wave numerical simulations of the measurements in a waveguide.}
\label{fig:fig6}
\end{figure*}

The metalens described in Fig.~\ref{fig:fig3} cannot be modelled in a waveguide due to the axial symmetry of the lens. Therefore, we manufacture a metalens with mirror symmetry along the $xz$-plane focusing reflected waves in the line parallel to the $x$-axis in the focal plane (Fig.~\ref{fig:fig4b}).
In the waveguide scenario we increase the number of the inclusions of the lens to 23 (see Supplementary Table~S2).
To analyze the metalens, we position the feed in the focal plane as shown in Fig.~\ref{fig:fig4d} (at a distance of $f=0.65\lambda=39$~mm). Based on the reciprocity principle, the metalens illuminated by an incident cylindrical wave from the feed should reflect a plane wave. Similarly to the previous case, the electric field distributions are plotted in Figs.~\ref{fig:fig6a}--\ref{fig:fig6f}. 
An incident cylindrical wave radiated from the focal point of the metalens (Figs.~\ref{fig:fig6a} and \ref{fig:fig6d}) is totally reflected with a plane wavefront (Figs.~\ref{fig:fig6c} and \ref{fig:fig6f}). The measured and simulated data confirm the desired performance of the metalens.
The above examples clearly demonstrate that arrays of small inclusions can fully reflect electromagnetic waves with any desired phase distribution.

\section*{Discussion}

Providing full control of reflected wavefronts, metamirrors are nearly transparent outside of the operational frequency band. This unique feature leads to a variety of exciting applications over the entire electromagnetic spectrum. At microwaves, being practically transparent at infrared and visible frequencies, metamirrors have a clear potential for breakthroughs in the design of antennas for various applications, in particular for satellites and for radio astronomy.
For example, the proposed layer can work as a large parabolic reflector for radio waves, while being deposited on a flat surface of a thin transparent film, dramatically simplifying deployment, or even on solar-cell panels of a satellite, not disturbing the panel operation. 
In radio astronomy as well as in satellite technologies, it will become possible to realize multi-frequency or multi-beam antennas using a parallel stack of such layers, each tuned to emulate parabolic reflectors at different frequencies and, if needed, with different focal points. This is possible due to the fact that outside of the design frequencies the layers are transparent and do not disturb the operation of the other layers.
Another exciting possibility is to exploit the extremely small focal distance of the proposed metalens, for instance, realizing extremely low-profile conformal antennas for mobile communications (the metasurface does not have to be flat and can be conformal to any object). 

The proposed concept may be scaled down to shorter wavelengths for a broad range of applications in nanophotonics and integrated optics. In particular, the proposed focusing metamirrors will open up new opportunities for micro- and nano-scale wavelength demultiplexing and nanosensing. Metalenses with the unprecedentedly small $f/D$ ratio provide huge concentration of optical energy that can be used for enhancing sensitivity of imaging and probing instruments at nanoscale.

\begin{methods}
\subsection{Metamirror fabrication.}
The inclusions of the metamirrors were manually manufactured of copper wires with 0.55~mm diameter and embedded in the supporting material. 
The first example metamirror consisted of 54 inclusions of six different types (described in Supplementary Table~S1) positioned periodically along the $y$-axis as shown in Fig.~\ref{fig:fig4a}. The distance between the adjacent inclusions was 14.14~mm. 
The metalens consisted of 23 inclusions (see Supplementary Table~S2) having mirror symmetry along the $x$-axis. The location of each inclusion is specified in Supplementary Table~S2. 

\subsection{Measurement characterization.}
A parallel-plate waveguide with the dimensions 80~cm $\times$ 90~cm (along the $y$ and $z$-axis, respectively) had the height of 14.1~mm and 15~mm for the cases of the anomalously reflecting metamirror and the metalens, respectively. The waveguide was fed with a stationary vertical coaxial feed. The inner conductor of the feed was connected to the bottom plate and the outer conductor to the top plate. The feed was positioned in the first and the second experimental setups at a distance of 80~cm and 39~mm from the metamirror, respectively. The manufactured one-dimensional metamirrors were positioned parallel to the edge of the waveguide at distances from the feed described in the previous section. The vertical coaxial probe positioned about 5~mm under the mesh measured near fields penetrated through the mesh. The 25~cm $\times$ 35~cm copper mesh had the period of 5~mm and the strip width of 1~mm. In the first experimental setup the mesh was located directly in front of the metamirror. In the second experimental setup the center of the mesh was positioned at a distance of 161~mm from the metalens. Microwave absorbing material blocks of 10~cm width were placed at the edges of the waveguide to reduce reflections. The stationary coaxial feed was connected to port 1 of Agilent Technologies' E8363A vector network analyzer, while the probe was connected to port 2. A field scanning system including the vector network analyzer and a PC connected to two moving platforms (along $y$- and $z$-directions) manufactured by Physik Instrumente was used to scan the field under the mesh in the horizontal plane. The scanning system measured transmission coefficient from port 1 to port 2 $S_{21}$ automatically with the step of 2.5~mm. The scanned area 30~cm $\times$ 24~cm (120 $\times$ 96 measurement points) was positioned in the center of the metal mesh. The spatial distribution of the transmission coefficient represents distribution of the $x$-component of the electric field.

\end{methods}

\begin{addendum}
 \item We thank Prof. N.~Engheta for useful discussions concerning scattering properties of the designed metamirrors. We additionally thank Mr.~M.~Albooyeh for help with post-processing of the simulated results.
 \item[Author contributions] S.A.T. and Y.R. conceived the concept and supervised the theoretical analysis and numerical simulations;
V.S.A. performed theoretical analysis, numerical simulations, device fabrication and measurements, and wrote the manuscript; J.V. and S.A.T. supervised the experimental measurements. All authors contributed to editing the manuscript.
 \item[Competing Interests] The authors declare that they have no
competing financial interests.
 \item[Correspondence] Correspondence and requests for materials
should be addressed to V.~S.~Asadchy (email: viktar.asadchy@aalto.fi).
\end{addendum}


\end{document}